\documentclass[a4paper,fleqn,usenatbib]{mnras}
\usepackage{mathptmx}
\usepackage[T1]{fontenc}
\usepackage{ae,aecompl}
\usepackage{subfigure}
\usepackage{graphicx}	
\usepackage{amsmath}	
\usepackage{amssymb}	

\title[Clustering in the Stellar Abundance Space]{Clustering in the Stellar Abundances Space}

\author[R. Boesso and H. J. Rocha-Pinto]
  {R. Boesso$^1$ and H. J. Rocha-Pinto$^1$\\
$^1$Observat\'orio do Valongo, Universidade Federal do Rio de Janeiro, Ladeira Pedro Ant\^onio 43,
20080-090, Rio de Janeiro, Brazil}

\date{Accepted 2017 October 17. Received 2017 October 17; in original form 2017 February 7}

\pubyear{2017}

\begin{document}
\label{firstpage}
\pagerange{\pageref{firstpage}--\pageref{lastpage}}
\maketitle

\begin{abstract}
We have studied the chemical enrichment history of the interstellar medium through an analysis of the $n$-dimensional stellar abundances space. This work is a non-parametric analysis of the stellar chemical abundance space. The main goal is to study the stars from their organization within this abundance space. Within this space, we seek to find clusters (in a statistical sense), that is, stars likely to share similar chemo-evolutionary history, using two methods: the hierarchical clustering and the principal component analysis. We analysed some selected abundance surveys available in the literature. For each sample, we labelled the group of stars according to its average abundance curve. In all samples, we identify the existence of a main enrichment pattern of the stars, which we call {\it chemical enrichment flow}. This flow is set by the structured and well defined mean rate at which the abundances of the interstellar medium increase, resulting from the mixture of the material ejected from the stars and stellar mass loss and interstellar medium gas. One of the main results of our analysis is the identification of subgroups of stars with peculiar chemistry. These stars are situated in regions outside of the enrichment flow in the abundance space. These peculiar stars show a mismatch in the enrichment rate of a few elements, such as Mg, Si, Sc and V, when compared to the mean enrichment rate of the other elements of the same stars. We believe the existence of these groups of stars with peculiar chemistry may be related to the accretion of planetary material onto stellar surfaces or may be due to production of the same chemical element by different nucleosynthetic sites.
\end{abstract}




\section{Introduction}

\indent The chemical abundances observed in the atmospheres of slightly evolved stars are fossil records of the chemistry of the gas cloud from which they were formed, allowing us to identify and study the Galactic chemical enrichment. Therefore, the evolutionary history of the Galaxy is preserved in the distribution of stellar abundances \citep{daSilva2012,bensby03,bensby05,bensby14}. On account of this, it is reasonable to expect that the stars observed today can be separated in similar chemical clusters of stars, which may reflect distinct clouds or galactic regions having the same enrichment history \citep{bland10}.

\indent Distinct chemical elements are formed in different astrophysical sites and time scales \citep{wallerstein97}. In principle, it should be possible to recover the formation and evolution history of the Galaxy from an analysis of the chemical abundances of stars born at different times. The chemical abundances in different regions of the interstellar medium reflect the integrated history of several stellar generations that enriched them. The distribution of stars in the chemical (elemental) abundance space \citep[hereafter $C$-space following][]{ting2012} can be used as a tool to recover this history. Stars are represented within the $C$-space by coordinates that correspond to its chemical abundances.

\indent Here, we perform a non-parametric analysis of the stellar chemical abundance space. The main goal is to study the structured organization of the stars in the abundance space. We seek to find groups of stars that have similar chemo-evolutionary history and setting these abundance clusters in an evolutionary sequence, similar to a taxonomic tree, on account on their chemical signatures \citep{FreemanBland2002}. A significant problem is that stellar abundance surveys only sample a subspace of the $C$-space, since there are no complete samples with abundance measurements for all chemical elements. Because of that, we analyse stars in an $n$-dimensional subspace, where $n$ is the number of elements in each sample, and our knowledge of the $C$-space will be limited by the dimension of this surveyed subspace.

\indent We used three statistical methods for multivariate data mining: hierarchical clustering, the principal component analysis (PCA) and the minimum spanning tree. They are pattern recognition techniques and are used here in order to hierarchically represent the $C$-space.

\indent To our knowledge, hierarchical clustering technique have been applied to stellar abundances only by \cite{daSilva2012,daSilva2015} . They investigated relations between kinematics, age and abundances for solar-type stars in the solar neighbourhood, and used dendrograms to divide the stars into groups based on the chemistry. There are very few studies in the literature that applies the PCA technique to stellar abundances. \cite{ting2012} analysed the stellar chemical abundances in various environments in our galaxy and in nearby dwarfs galaxies, and represented these abundances in a new space with dimension smaller than that of the original components, and related to nucleosynthetic processes; and  \cite{andrews2012} made an investigation on the chemical abundances of bulge stars through the PCA technique in order to study their formation and evolution. There is a recent study about chemical tagging, \cite{hogg2016}, which seeks similar stars in the velocity and chemical space.

\indent Large surveys of chemical abundance were published in recent years, establishing observational constraints to the Galactic chemical evolution. Nine of these surveys were chosen to our analysis, with large numbers of stars and large amount of elements: \citet{edv93}; \citet{fulbright2000}; \citet{gratton2003}; \citet{reddy03}; \citet{reddy2006}; \citet{takeda2008}; \citet{neves2009} and \citet{adibekyan2012}. Each sample covers a different range of metallicity and different Galactic populations.

\indent This papers presents a pilot study of the structured organization of stars in the chemical abundance space. We are interested in finding clustering of stars that have peculiar (non-standard) chemistry when compared to the main chemical enrichment flow of the Galaxy. We also study the hierarchical relation of these peculiar stars with respect to other clustering of stars. We demonstrate that similar results are obtained from distinct sets of stellar abundances in samples published by different authors. Our next step will be to investigate in a similar way more complete and representative stellar samples now available from the many ongoing large spectroscopic stellar surveys, e.g., Gaia-ESO \citep{gilmore2012,RandichGilmore2013} and APOGEE \citep{Majewski2015}.

\indent This paper is organized as follows: in Section 3 2 we explain how the chemical evolution formalism can be used to advance expected properties of the $C$-space. In Section 3 a detailed description of the samples used. In Section 4 we describe the methodologies used, the hierarchical clustering technique and the principal component analysis, and their application to our samples. Section 5 presents the results and discussions. Finally, Section 6 gives a summary of the main conclusions.

\section{Abundance space in light of the theory of chemical evolution}\label{chemical_evolution}

In this Section, we explain some concepts needed as background to understand what are the expectations for the behaviour of the stars in the $C$-space. Stars are not be found scattered throughout the $C$-space. They should instead follow well structured and hierarchical paths which depend on the history of chemical enrichment up to the point at which they were formed.   

The equations that describes the abundance evolution of an element in the interstellar medium can be found in \cite{matteucci03} or \cite{tinsley80}. It is possible to show that the relation between the increase in the abundances $X_i$ and $X_j$ of two elements $i$ and $j$ is:

\begin{equation}
 \frac{dX_{i}}{dX_{j}}=\frac{(R_i-X_{i}R)+y_{i}(1-R)}{(R_j-X_{j}R)+y_{j}(1-R)}\equiv\frac{y_{{\rm eff},i}}{y_{{\rm eff},j}};
 \label{eq_1}
\end{equation}

\noindent where $R$ is the gas mass return fraction, $R_i$ and $R_j$ are return fraction of mass in the form of $X_i$ and $X_j$, respectively. In practical terms, the effective yield $y_{{\rm eff},i}$ of $X_i$ is variable over time, and given by:

\begin{equation}
 y_{{\rm eff},i}=\frac{(R_i-X_iR)}{(1-R)}+y_i,
 \label{eq_2}
\end{equation}

\noindent where $y_i$ is the net yield of elements $i$ newly produced and ejected.

\indent The terms in the equation \ref{eq_1} are similar to the the classical equation for the metallicity evolution in the Simple Model of Galactic Chemical Evolution \citep{schmidt63,tinsley80,pagel97,matteucci03}, except that in this equation there is still no assumptions regarding \textquotedblleft instantaneous recycling approximation\textquotedblright. When the instantaneous recycling approximation is used, the relation between the increase in the abundances of two elements $i$ and $j$ can be rewritten as:

\begin{equation}
 \frac{dX_{i}}{dX_{j}}=\frac{y_{i}}{y_{j}};
 \label{eq_3}
\end{equation}

\indent When comparing chemical elements that do not contribute to the production and depletion of one another, Equation \ref{eq_3} can be taken as a linear relation or a simple correlation between $X_i$ and $X_j$. Out of the instantaneous recycling approximation (see Equation \ref{eq_1}), the relation between $X_i$ and $X_j$ can still be interpreted as a local linear regression having a variable correlation coefficient due to the difference in the average lifetime of stars that contribute to production of $X_i$ and $X_j$.

\subsection{Stars in the $C$-space}

\indent In the $C$-space, each star is represented by a point whose coordinates are ($X_1$, $X_2$, ... $X_n$), where each $X_i$ represents the abundance of the $i$th chemical element. The distribution of stars in this $n$-dimensional space should reflect the relations of Equation \ref{eq_1}. Thus, each relation between the abundance of two elements ($X_i\times X_j$), that may be represented in a plot (as usual in most works dealing with chemical abundances), corresponds to the projection of this $n$-dimensional distribution in the abundance subspace $X_iX_j$.

\indent The correlation between $X_i$ and $X_j$ is directly related to $dX_i/dX_j$, which is equivalent to the ratio of the effective yields. Since this argument is valid for any pair $(i,j)$, we can construct a $n\times n$ correlation matrix that summarizes the concomitant variations in this $n$-variational space. It is clear that the elements $a_{ij}$ of this matrix are proportional to $\langle dX_i/dX_j\rangle$, since the derivative corresponds to the slope of a linear relation between $X_i$ and $X_j$. It is necessary to consider the proportionality to the average $\langle dX_i/dX_j\rangle$ because the ratio of yields may change with time.

\indent  This discussion aims to introduce a concept that will be very important for the interpretation of the clustering of stars that we will find in the chemical abundances space: there is a straight correspondence in the occupation of this space and in the correlations that can be observed between pairs of elements abundance $(X_i,X_j)$. The most trivial case is that in which the instantaneous recycling approximation is valid for all elements considered and $y_i/y_j=1$, $\forall$  $i,j$. In this case, the $n\times n$ correlation matrix would have all elements equal to 1 and $n$-dimensional space would be populated only along its hyperdiagonal\footnote{This line is also often called the space diagonal.}, i.e., the $n$-dimensional analogue of the line $x = y$ in the two-dimensional space.

\indent If $y_i/y_j\neq1$, $\forall$ $i,j$, but constant and non-zero at all times, the relation between the abundances $X_i$ and $X_j$ will be still linear although it would no longer occupy the diagonal of $X_iX_j$ plan. In the $n$-dimensional abundance space, the stars would be found in an $n$-dimensional line, having projections onto the $X_iX_j$ plans corresponding to linear relations with an angular coefficient $y_i/y_j$. We can use the sample correlation matrix to foresee how should be the occupation of the abundances space from its various bi-dimensional projections. The narrower and more well-defined are the relations $X_i\times X_j$, $\forall$ $i,j$, in a sample, the more confined to an $n$-dimensional line are the stars of this sample. This $n$-dimensional line has the same role of the principal axis of the $C$-space, along of which is the maximum variance of the data.

\subsection{Chemical enrichment flow in the $C$-space}

\indent We will call the gain in metals which occurs along this $n$-dimensional line, or principal axis, as {\it chemical enrichment flow}. Along this enrichment flow, the relation between increments of abundance in the $n$ elements is constant or nearly constant, so that a star 10 times poorer than the Sun in Fe is also about 10 times poorer than the Sun in its other heavy elements. That is, the relative abundance pattern of these stars varies little or nothing.
  
\indent The chemical enrichment flow sets the pace at which the average abundances of the interstellar medium grow. Any classification and clustering method that is applied to the abundances space should evidence it as a zero-order trend. Because it is an $n$-dimensional line, extending from the region of the smallest to the largest abundances, for all elements treated here, the statistically segregated chemical clustering of stars are distributed, at first sight, along this line, leading to the seemingly trivial finding that such partition can be interpreted as formed by metal-poor, intermediate metallicity and metal-rich stellar groups \citep[e.g., see][and Section 5 of this work]{daSilva2012}.

\indent We can remove this zero-order trend from the data before using them in classification methods. This can be done by using the abundance ratio space, [$X$/Fe], instead of the abundance space, [$X$/H]. We choose not to perform the analysis within the abundance ratio space for the following reasons:

\begin{enumerate}
\item the analysis of chemical clustering of stars is already performed in the abundance ratio space, although not explicitly, in several studies that do chemical tagging \citep{FreemanBland2002} in plots [$X$/Fe] vs. [Fe/H];
\item our intention is to represent the chemical clustering of stars in an evolutionary hierarchy (hence the use of classification trees). For this we need to implicitly have the variable time in our data, which is possible with the use of abundances [$X$/H].
\end{enumerate}

\subsection{Stars with peculiar chemistry}\label{stars_peculiar_chemistry}

\indent We are interested in finding clustering of stars with peculiar chemistry, which are situated in regions out of the enrichment flow in the abundance space.

\begin{figure*}
\includegraphics[width=98mm]{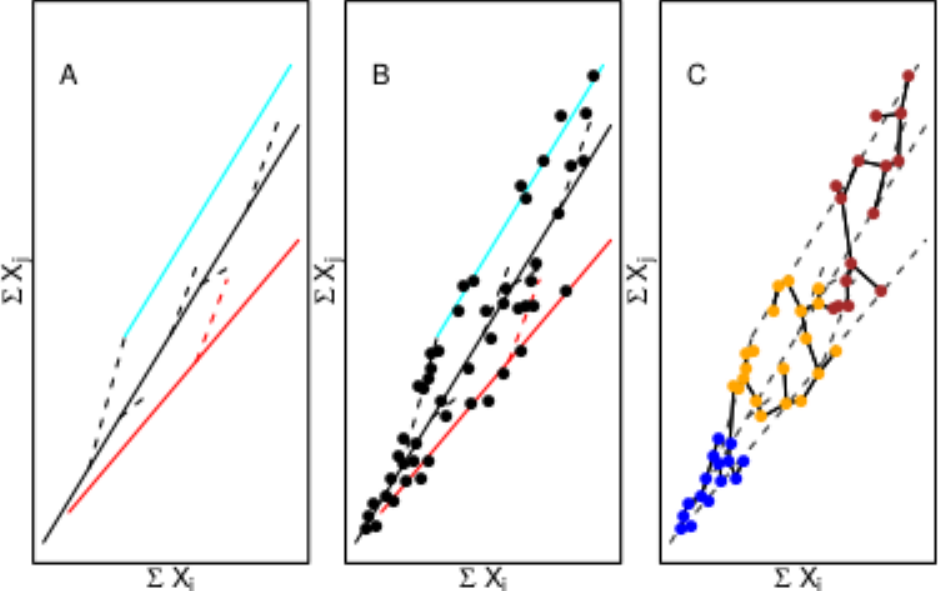}
\caption{Schematic representation of a stellar chemical abundance subspace. This figure is a simple and theoretical scheme of how stars must be distributed within the chemical abundance space. Each axis of theses panels represents a particular subspace of the $n$-dimensional abundances space. Panel A: The chemical enrichment flow, that dictates the pace at which the  chemical abundances grow, is represented by the solid black line. Small dashed lines which are pronounced from the chemical enrichment flow towards increasing abundances are called effluents branches. The extent of these dotted lines depends on the efficiency of the mixing process and chemical homogenization of the interstellar medium, because it depends on the relative isolation of that part of the galactic gas. If this isolation is maintained for a long time, the galactic region can inherit an entirely different chemical pattern (cyan solid line). Another case is that for which a galactic region has an initial mass function different from that taken as an average of the galactic disk and this region will be enriched by its own chemical enrichment flow (red solid line). Panel B: Example of a typical stellar sample containing stars formed along these various enrichment lineages of this abundances space. Panel C: The same sample in Panel B, now separated into groups (clustering of stars, represented from different colours), depending on the chemical abundance, identified by the hierarchical clustering method and linked by a minimum spanning tree.}
\label{esquema}
\end{figure*}

\indent The existence of stars with peculiar chemistry can be understood in light of the Equation \ref{eq_1} where the ratio of effective yields for a certain number of chemical species is substantially different from that ratio which characterizes the chemical enrichment flow at the same time $t$. According to the equation, the effective yield would change when the proportion of stellar masses which die in a time $t$ varies\footnote{For example, when the stellar lifetime is higher than the typical scale of star formation, the nucleosynthetical products of this star are delivered to the interstellar medium in a \textquotedblleft delayed\textquotedblright \ way. This delayed production is characteristic of SN Ia and AGB due to the non-negligible lifetime of the progenitor stars.}, either on account of the finite stellar lifetimes or of a varying initial mass function.

\indent Stars born in these regions would show peculiar abundance ratios. The abundances space is sketched in Figure \ref{esquema}A. Each axis of this figure schematically represents a particular subspace of the $n$-dimensional abundance space. In these subspaces, the stars should be distributed mainly along an $n$-dimensional line, represented by the solid black line, which we called \textquotedblleft chemical enrichment flow\textquotedblright. According to the equations derived above, when the ratio of the effective yields of elements $i$ and $j$ does not vary over time, the abundance subspace $X_iX_j$ is populated only along a single line. Later variations in the effective yields of a few pairs $(X_i,X_j)$ should result in effluent branches (represented by the small dashed lines that depart from the chemical enrichment flow toward increasing abundances). 

\indent The extension of these branches depends on the efficiency of mixing process in the interstellar medium, because it requires the relative isolation of that portion of the galactic gas, such that this peculiar chemical pattern created from a partial enrichment (i.e., by a few specific stellar masses) is not obliterated by further mixing of ejecta from different masses. If this isolation is maintained for a long time, before the ratio of specific yields approach a constant value, this galactic region can inherit an entirely different chemical pattern (represented by the cyan solid line), although this chemical pattern can subsequently grow in a similar rate to the main flow. That is, the enrichment pace (average chemical enrichment due to the combined mixing of ejecta from several stellar masses exploding at a given time) was stabilized, but only after the characterization of a distinct initial chemical pattern. Another case is one in which a galactic region has an initial mass function different from the one which can be taken as an average of the galactic disk; in this case, this zone is enriched following a different chemical enrichment flow (solid red line). Finally, each of the populated regions of the abundance space may have its effluent branches, due to the peculiar chemical patterns imposed on an average chemical pattern.

\indent This qualitative discussion aims to show why the abundance space should not be randomly populated by stars, but highly structured.

\indent A typical sample of this stellar abundance space should contain stars along these various enrichment \textquotedblleft lineages\textquotedblright, as shown in Figure \ref{esquema}B. Ideally, as we propose to classify the stars using their abundances, in a fashion similar to the taxonomic concepts of biology, we want to be able to detect the hierarchical lineages in the enrichment of the galactic disk. Our success could be described in how closely we can reproduce the hierarchical lineages, effluent branches, etc., to which each star belongs, as well as be able to date these enrichment events. However, we must pay attention to the limitation of these techniques. The methods we use are unable to separate the enrichment lineages if the separation between them is not as large as the total variation of the abundances along the average chemical enrichment flow (represented by the black line in Figure \ref{esquema}A). The consequence of this is that the maximum variance along this flow dominates over the other variances (along other lines of this $n$-dimensional space), so that the main abundance clusters to be found are trivially interpretable as those of metal-poor, metal-intermediate  and metal-rich stars, and  each of these may also have similar divisions. Figure \ref{esquema}C illustrates this. 

\begin{table*}
\begin{minipage}{170mm}
 \caption{Samples analysed in this work.}
 \label{sample}
 \centering
 \begin{tabular}{@{}ccccc}
  \hline
  Survey & no. of Stars & no. of Chemical Elements & type of stars & Galactic Coverage\\[2pt]
  \hline
  Edvardsson et al. (1993) & 189 & 13 & F and G dwarfs stars & disk (thin, thick) \\[2pt]
  Fulbright (2000) & 168 & 14 & K stars & disk (thin, thick), halo \\[2pt]
  Gratton et al. (2003) & 150 & 13 & FGK sudwarfs and subgiants  & disk (thin, thick), halo \\[2pt]
  Reddy et al. (2003) & 181 & 27 & F and G dwarfs stars & disk (thin, thick) \\[2pt]
  Reddy et al. (2006) & 176 & 22 & F and G dwarfs stars & disk (thin, thick), halo \\[2pt]
  Takeda et al. (2008) & 322 & 18 & G giants stars & thin disk \\[2pt]
  Neves et al. (2009) & 451 & 13 & FGK dwarfs stars & disk (thin, thick) \\[1pt]
  Adibekyan et al. (2012) & 1111 & 13 & FGK dwarfs stars & disk (thin, thick), halo \\
  \hline
 \end{tabular}
\end{minipage}
\end{table*}

\indent In Figure \ref{esquema}C we show the same sample of Figure \ref{esquema}B, now separated into clusters of stars, which are formed by chemical similarity by the method of hierarchical clustering (see Section \ref{methodology}) and connected by a minimum spanning tree. Note how the actual a priori hierarchy in Figure \ref{esquema}B is obliterated precisely along the main flow of chemical enrichment. Stars of different lineages (the lines of different colours in Figure \ref{esquema}B) can be taken as a group just based on the average value of their of abundances in those variables that explain the maximum variance of the $C$-space. However, the minimum spanning tree, Figure \ref{esquema}C, is still able to join several stars coming from a common lineage. We hope to be able to find real clusters as well as chemically peculiar subgroups {clusters of stars with a lower number of stars} as subsets of the larger groups that merely divide the abundance space along the chemical enrichment flow.

\indent It is worth noting, as can be inferred from Figure \ref{esquema}C, that a metal-rich star of the thin disk is chemically closer to a metal-rich star of the thick disk than to a metal-poor star of the thin disk itself. The methods of hierarchical clustering applied in this work find stars with similar chemistry in [$X$/H]. They are less sensitive to [$X$/Fe] and so are not used to recover or identify components of galactic structures such as the thin and thick disks.

\section{The database}

We have selected samples on the literature based on their coverage of the abundance space, resulting in 8 surveys. The selection criterion we use to define the sample coverage was the number of surveyed stars and the number of elements in the survey, simultaneously: only those surveys having more than 150 stars and 10 elements were used. In Table \ref{sample}, we list the samples used in this study.

\begin{figure}
\includegraphics[width=90mm,trim=0cm 0.7cm 0cm 1cm,clip=True]{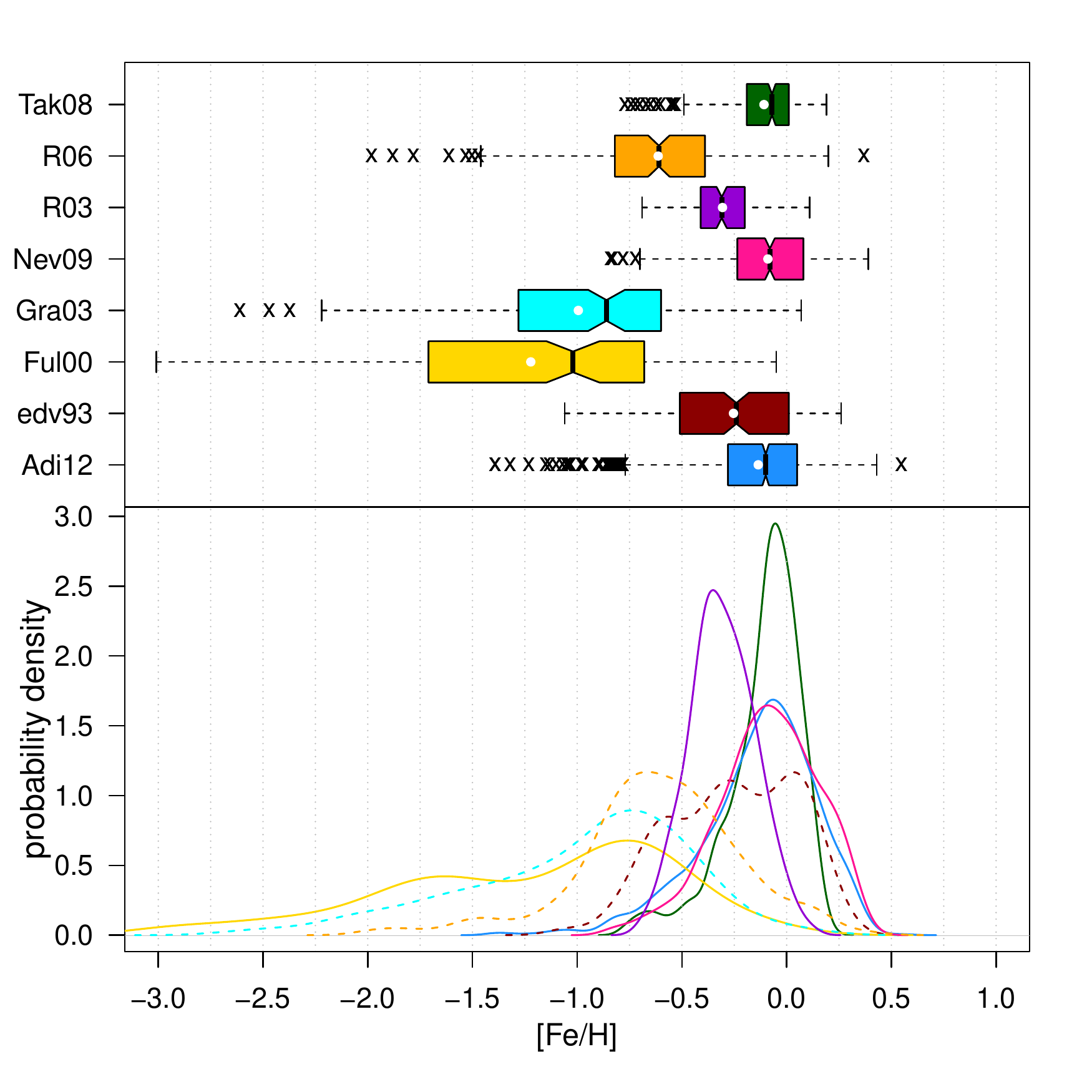}
\caption{Metallicity distribution for surveys that constitute our sample. In the top panel, the distributions are shown by box diagrams while the bottom panel the same distributions are represented by a probability density function smoothed by gaussian  kernel. The samples of this figure are differentiated according to the following colours: \citet[Adi12]{adibekyan2012}, blue; \citet[Nev09]{neves2009}, pink; \citet[Tak08]{takeda2008}, dark green; \citet[R06]{reddy2006}, orange; \citet[R03]{reddy03}, purple; \citet[Gra03]{gratton2003}, cyan; \citet[Ful00]{fulbright2000}, yellow; and \citet[edv93]{edv93}, dark red. In both panels, and the top panel they are represented by abbreviations.}
\label{distribuicao_abundancia}
\end{figure}

\indent Each of the samples was built with a particular goal by the respective authors. Taken together, they form a very heterogeneous sample that unequally covers the space of the chemical abundances. For example, some of these surveys aimed to study metal-poor stars \citep{fulbright2000}, and others, metal-rich stars \citep{neves2009}. This focus on different parts of the abundance space is more clearly highlighted by the [Fe/H] distribution of each survey, as shown in Figure \ref{distribuicao_abundancia}. The abundances coverage varies,  not only from a survey to another but also from one element to another.

\indent Such heterogeneity of the data may limit the statistical significance of our analysis, since the unequal coverage of the abundance space, aggravated by the diversity of abundance measurement methods and sample selection from one survey to another, destroys the real proportion in which chemically different stellar groups in the solar neighbourhood are represented in this study.

\indent This means we can not use a combined sample of these individual surveys to study the distribution of the abundance of an element in the solar neighbourhood. Instead, we are interested in finding chemically distinct and extreme groups. Our work do not seek to find the proportion of these stars in the solar neighbourhood, but attest their existence, incorporating them into a hierarchical classification based on the chemical properties.

\indent We can illustrate this sample heterogeneity problem in our database with that found by a biologist who seek to taxonomically classify species based on small number of species surveyed heterogeneously. Eventually, the biologist might be unable to find a family or a genus not present in his sample, but still should be able to successfully group the species in phyla, kingdoms and orders. Our expectation, therefore, is not to provide a full inventory of the stellar content of the solar neighbourhood, but identify and hierarchize the principal chemical classes in which the stars are distributed.

\indent Most of our samples, with one exception \citep{neves2009}, do not contain a complete data set, that means their sample do not have abundance measures for all elements for all stars. As the method applied by us does not allow missing values, since the analysis is developed by comparing the values between objects, we applied some cuts, discarding stars or elements in order to have full complete samples.

\section{Methodology}\label{methodology}

\indent In our study we make use of three different methodologies. Since the goal of this work is to find clustering of stars in the abundance space and study their enrichment pattern, we apply three methods: the hierarchical clustering technique, the principal component analysis (PCA) and the minimum spanning tree. These are methods of unsupervised pattern recognition \citep{Everitt2001,GorbanZinovyev2009,hartmann2015}.

\indent An unsupervised analysis occurs when in the input data set there is no information about the class associated with each example. In the case of the hierarchical clustering technique, the only information we have is the dissimilarity matrix.

\indent To assist in the following explanation, we assume that the data set for m objects in an n-dimensional space can be described as follows: $X^{(i)}=(X_{1}^{(i)},X_{2}^{(i)},X_{3}^{(i)},...,X_{n}^{(i)})$, where $ i = 0,1,2, ..., m-1 $, such that $m$ is the number of objects and $n$ is the number of variables. In this work, stars are the objects and the chemical abundances for each element are the variables of our data set.

\subsection{Hierarchical clustering technique}

\indent The hierarchical clustering aims to join objects, so that there is homogeneity within a group but heterogeneity among groups. Initially, we consider that the number of groups is equal to $m$, which is the number of objects in the data set. In this way we have that the set of groups can be written as: $C = {c_{1},c_{2},,c_{3},...,c_{m}}$, where $c_{i} = \{X^{(i)}\}$. The distance between two groups is the distance between the variables of two objects. We define the distance between the group $c_{i}$ and $c_{j}$ by the distance function $d_{c}(i,j)$, which is the distance between two points. We use euclidean distance, which can be described analytically by the equation:
\begin{equation}
d_{c}(i,j)=\left[\sum_{a=1}^{n}(X_{a}^{i}-X_{a}^{j})^{2}\right]^{\frac{1}{2}},
\label{eq1}
\end{equation}

\noindent where $X_{a}^{i}$ is the value of the variable $a$ for the object (or group, in the case $c_{a}^{i}$) $i$ e $X_{a}^{j}$ is the value of the variable $a$ for the object (or group, in the case $c_{a}^{j}$) $j$.

\indent By recursive clustering of the data by similarity, based on the distance between them, the number of groups progressively decreases, so that the groups with lesser distance value are combined, forming new groups. The groups form in pairs, agglomeratively and hierarchically. The process of forming subgroups and groups stops after $m-1$ interactions, where $m$ is the number of objects in the data set. At this point, there is a single group containing all objects.

\indent For each iteration that occurs, forming new groups, a dissimilarity matrix is created. A dissimilarity matrix is built from the measured distances between the location of the objects in $C$-space. In the first iteration, where each object in the data set is considered a group, the dissimilarity matrix is an array of distances between an object $i$ and another object $j$. After forming the first group, from the two objects with the lowest distance value, we do not only have groups with one member per group. The new group has two members. Therefore, a new matrix of dissimilarity is created. The method for determining the distance between the groups used was the average linkage, whose distance $d_{c}(k,l)$ from the new group $c_{k}$ to another group $c_{l}$ is calculated by the average distance between the objects of the first group and the objects of the second group and can be described by the equation:
\begin{equation}
d_{c}(k,l)=\frac{1}{|c_{k}||c_{l}|}\sum_{i\in c_{k},j\in c_{l}}d(i,j),
\label{eq2}
\end{equation}

\noindent where $c_{k}$ e $c_{l}$ are the numbers of objects in the groups $k$ e $l$, respectively. We have chosen the average linkage method because other methods (such as complete linkage and single linkage)\footnote{The distance between groups in the complete linkage corresponds to the greatest distance between two objects of different groups and in the single linkage corresponds to the smallest distance between two objects of different groups.} are more sensitive to abundance outliers.

\indent For the application of the technique, it is necessary that the data is in a suitable format for the analysis. The completeness of the data is essential, such that objects that do not have some of the measured properties should be discarded previously to the analysis. During the data preparation, we excluded these objects in order to not significantly reduce the size of our final sample.

\indent We used dendrograms to represent the results. The dendrogram is a graphical tool which can objectively illustrate the hierarchy of the clusters.

\subsection{Principal component analysis}

\indent Principal component analysis (PCA) is a suitable technique for the interpretation and description of multivariate data, by allowing the description of a complex data set through the reduction of its dimensionality. PCA applications in chemical abundance analysis have been carried out by \cite{ting2012} and \cite{andrews2012}.

\indent When we study a data set, with a large number of objects $m$ and variables $n$, the existence of structures and correlations between the variables facilitates the understanding of the data. Therefore, an alternative is to work not with variables, which are many, but with derived variables.

\indent PCA performs a transformation, creating a new subspace whose number of components is smaller than the number of variables, showing which variables in a data set are the most significant and how these variables are correlated. This analysis allows the representation of the set of variables (abundances) by a set of orthogonal vectors in a $C$-space (chemical space), so that these vectors or components are able to explain the maximum variation of the data using the smallest possible number of components \citep{pearson1901}.

\indent The components originated from the transformation of $n$-dimensional space into a smaller number of variables can be described according to the generalized equation below:
\begin{equation}
PCk = \beta_{k}X = \beta_{k1}X_{1}+\beta_{k2}X_{2}+...+\beta_{kn}X_{n}=\sum_{a=1}^{n}\beta_{ka}X_{a},
\end{equation}

\noindent where $\beta_{k}$ is a vector of $n$ constants, $PCk$ represents the $k$-th principal component and the vectors ($\beta_{1}$, $\beta_{2}$, $\beta_{3}$, ..., $\beta_{n}$) uncorrelated.

\subsection{Minimum spanning tree}

\indent The minimum spanning tree is a graphical data clustering technique that consists of connecting a set of objects, which we call tree nodes, into pairs of objects through edges, without closed loops, so that the total length of the tree is minimum. The length of the tree is the linear sum of the $l$ lengths of the edges of the tree. The length of each edge corresponds to the distance between two objects $i$ and $j$ connected, which can be represented by the euclidean distance, described in the equation \ref{eq1}.

\indent According to \cite{barrow85}, each node $i$ in the tree corresponds to an object $X^{i}$ of the data set. A node $i$ of the data set is chosen arbitrarily and a smaller edge connects this node to another node $j$. This process is performed continuously until all nodes are connected to at least one other node, so that all objects are part of the tree and without any closed loop of nodes. In each process of joining the nodes (represent the objects) by an edge (represent the distance between objects) a subtree is formed. The minimum spanning tree selects, within $n(n-1)/2$ possible different separations, a subset representing the similarity (smallest distance) between objects. At the end of the connections, we have a tree with all nodes connected, with the smallest possible length, forming the minimum spanning tree.

\indent The minimum spanning tree can be divided into smaller groups in order to find similar isolated groups. In this case, a node $i$ is connected to another node $j$ by an edge ${i,j}$ if the distance between objects is smaller than a given threshold $\theta$, so that $d(i,j)<\theta$. Small values of $\theta$ provide a setting in subtrees, where each tree represents a group (see Figure \ref{esquema}). Many large values of $\theta$ can lead to a configuration where only one group is found, forming a large single tree.

\indent As this technique generates a tree of smaller length, it is expected that it will satisfactorily represent patterns of clustering of stars within the space of chemical abundances.

\indent In this paper, the points are represented by the stars and the separation between them, the edges, is represented by the distance in stellar abundance. We generated minimum spanning trees for the first two principal components (first principal component, PC1, and second principal component, PC2), in order to study the chemical patterns in our samples.

\section{Results}

\indent In this Section, we discuss the results obtained and describe relevant details that are observed regularly in all samples.

\indent This analysis is a pilot study, a first analysis and demonstration of the methodology applied here. We select multiple surveys in the literature in order to test the methodology and compare the results for the different surveys, and to make sure that the results are consistent and independent of the survey used. The surveys used here are not similar, as can be seen in the metallicities distribution of Figure \ref{distribuicao_abundancia}. Once the consistency of the results is established independently of the survey used in the analysis, we will later apply the methodology used here to large surveys with a large number of stars and chemical abundances for different elements.

\indent Besides finding groups of stars having similar chemical enrichment, we seek to obtain, within these, subgroups that did not follow the \textquotedblleft typical\textquotedblright chemical enrichment history; that is, stars particularly super enriched by a particular element compared to the average enrichment of other elements.

\begin{figure}
\centering
\includegraphics[width=90mm]{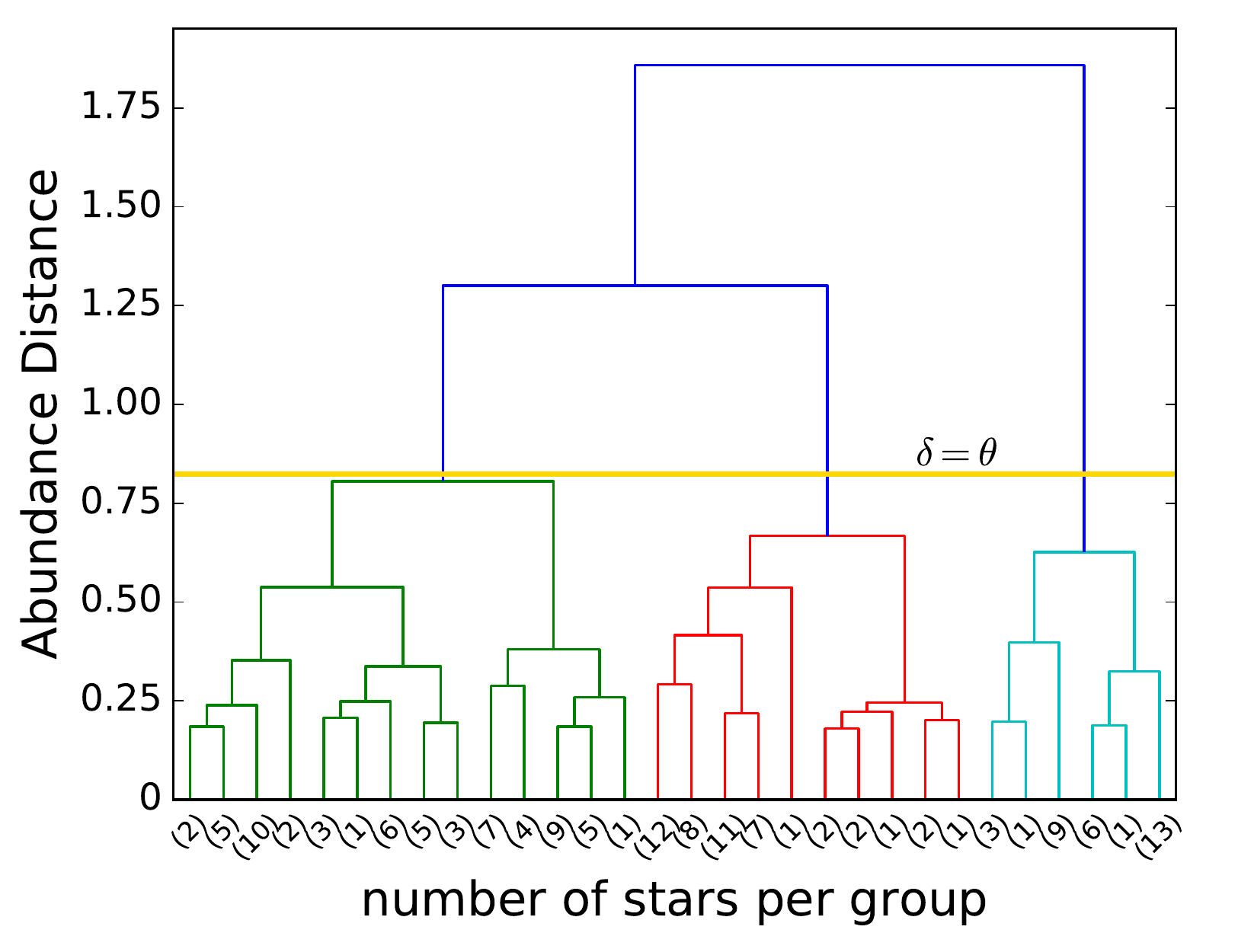}
\caption{Dendrogram representing the hierarchical clustering of the sample of \citet{edv93}. In the y-axis, we list the distance between the object and the centre of each group, whereas the x-axis has as output the ramifications of the groups with the number of objects in parentheses. There are 3 main groups according to the cut-off criteria (explained in Section \ref{HC} and from Figure \ref{decaimento_edv3}). The yellow line represents the cut in the dendrogram for $\delta = \theta$, where $\delta$ represents the height on the y-axis and $\theta$ the threshold value according cut-off criteria. Dendrograms similar to this were constructed for the other samples, but are not shown here on account of limiting the paper size.}
\label{tree_edv93}
\end{figure}

\begin{figure}
\centering
\includegraphics[width=90mm]{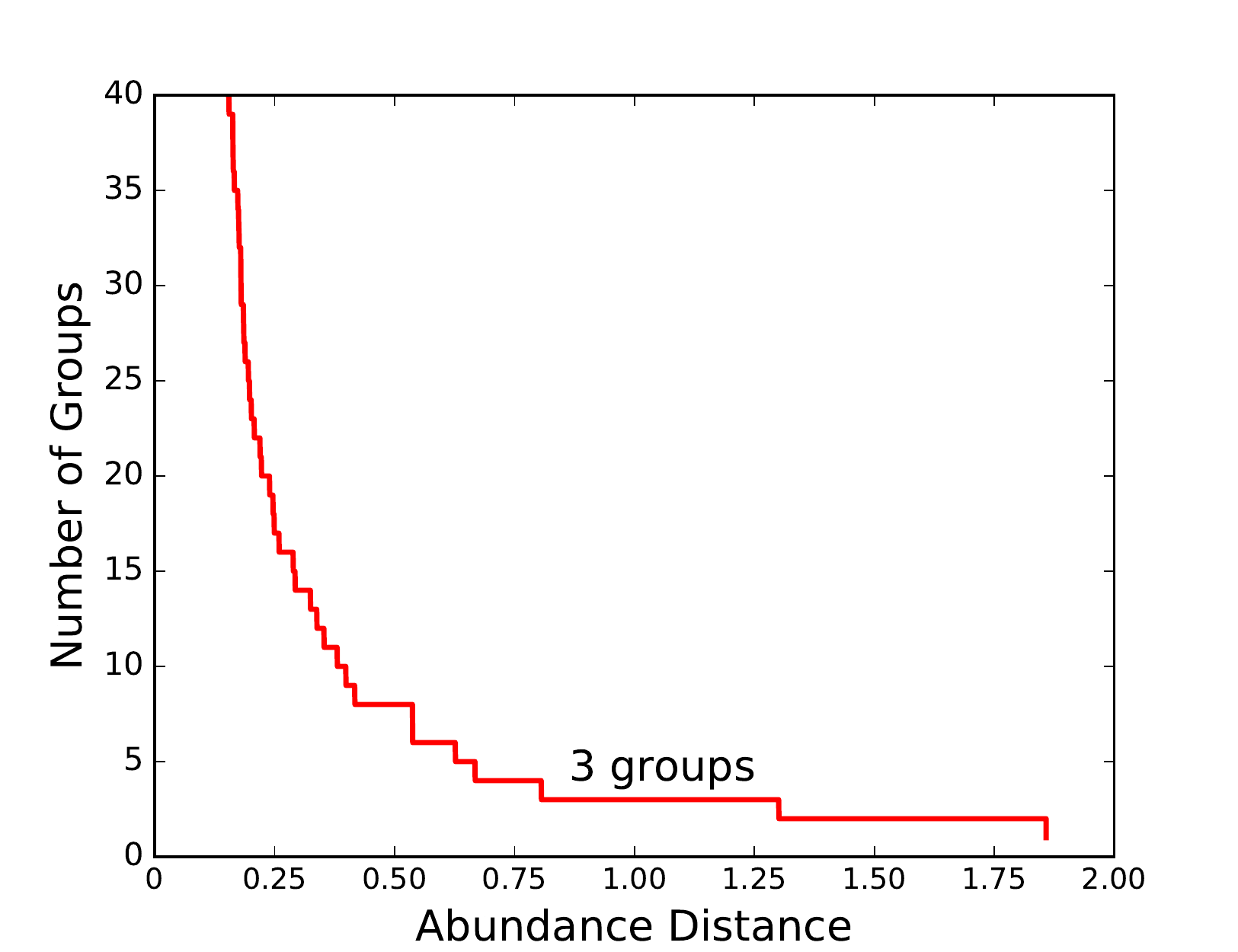}
\caption{ Decay curve of the number of groups for \citet{edv93} sample, used to establish the cut-off criteria. The curve shows the number of groups obtained from the hierarchical clustering technique (shown in Figure \ref{tree_edv93}) as a function of the abundance distance. According to the cut-off criteria, the cut is performed in a height on the y-axis of the dendrogram where the number of groups remains constant over a greater range of abundance distance. The number of groups according to the cut-off criteria is 3 for \citet{edv93} sample.}
\label{decaimento_edv3}
\end{figure}

\indent We will analyse and discuss the abundance patterns in these groups and subgroups according to the nucleosynthesis processes responsible for the variation in the average abundance. The data used in both methods (hierarchical clustering and PCA) are the ratio of the abundance by number of element $X$ to hydrogen with respect to the Sun ([$X$/H])\footnote{[$X$/H] $= \log$ [n($X$)/n(H)]$_{\rm\star}-\log$ [n($X$)/n(H)]$_{\rm\odot}$}.

\subsection{Hierarchical Classification and the Chemical Abundance Curves}\label{HC}

\indent The technique of hierarchical clustering was applied to each sample listed in Table \ref{sample}. As described, the stars are clustered according to similar abundances. The results are presented in a dendrogram (e.g., see Figure \ref{tree_edv93}). In a dendrogram, the y-axis gives the distance in the abundance space between the object and the centre of each group, and the x-axis has the objects as output. The distance is measured in the unit of data, in our case, dex.

\begin{figure}
\centering
\includegraphics[width=90mm, trim=0.6cm 0.2cm 1cm 1cm,clip=True]{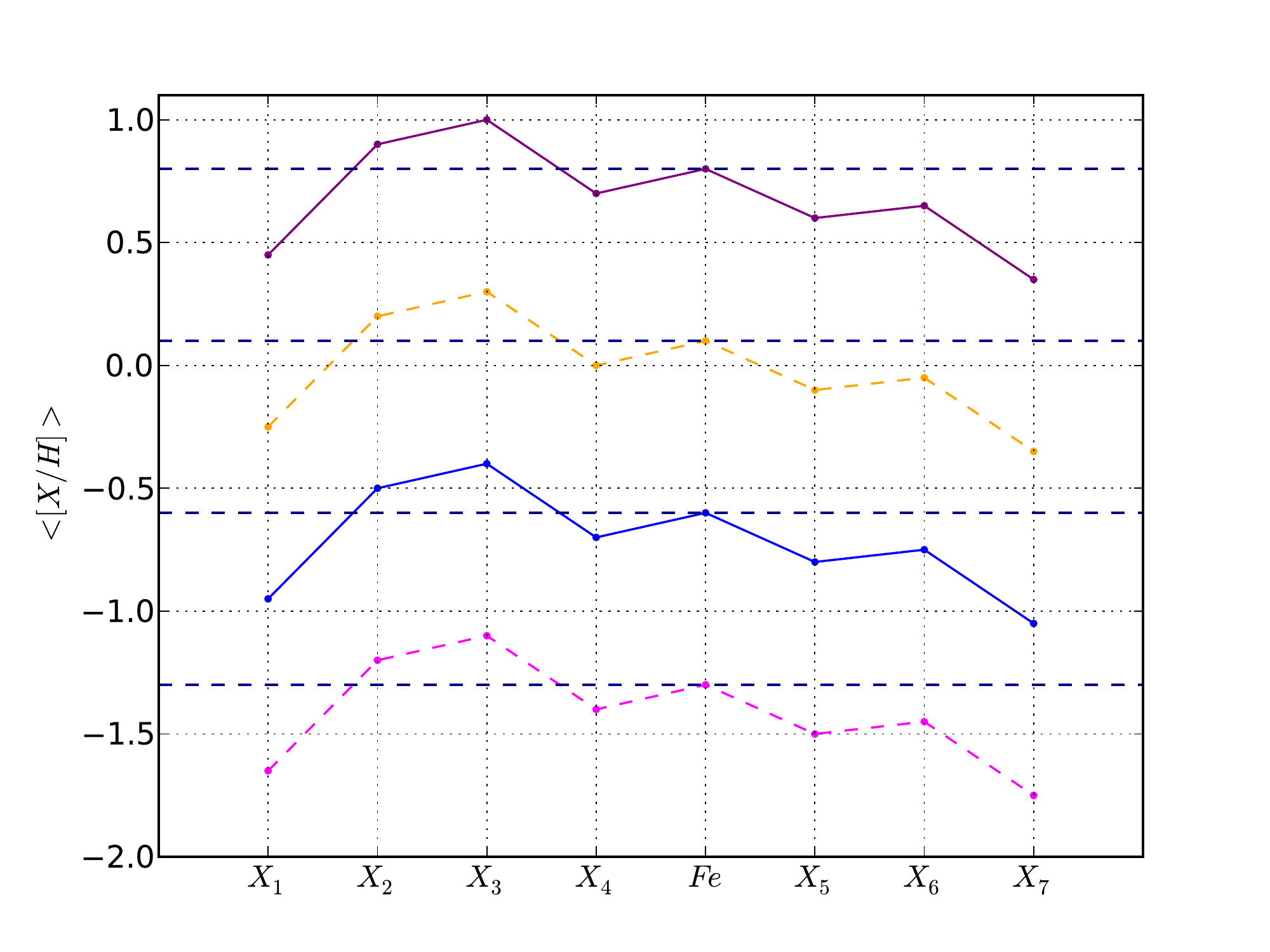}\\
\includegraphics[width=90mm, trim=0.6cm 0.2cm 1cm 1cm,clip=True]{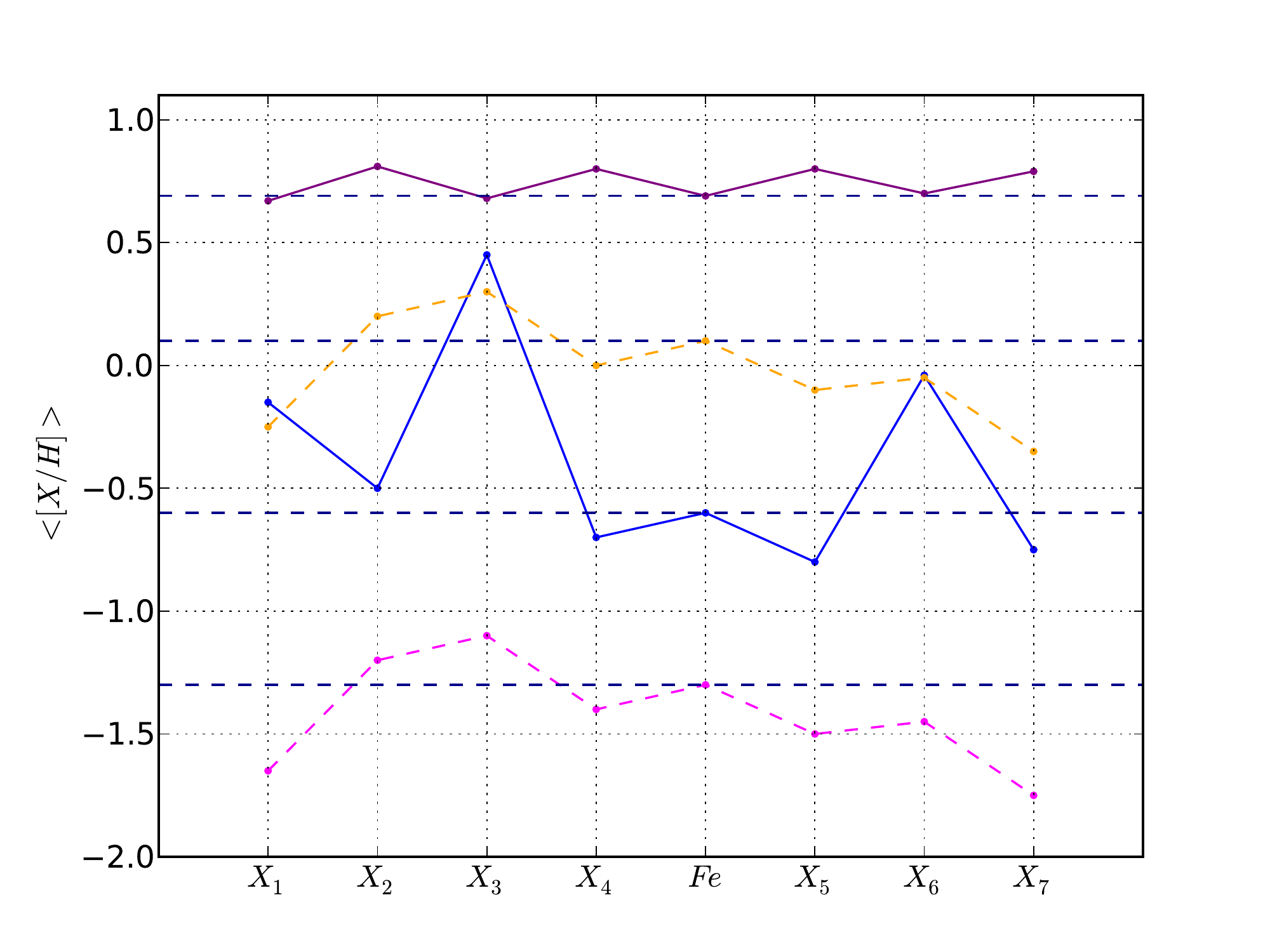}
\caption{Theoretical representations of average abundance curves for different types of groups. Top: representation of average abundance curves for groups of stars born from an interstellar medium having nearly constant average enrichment rate for all elements. Bottom: representation of average abundance curves for, besides 2 groups that have proportional chemical abundance patterns (magenta and yellow curves in the bottom plot),  a group of stars born from an interstellar medium with large variation for the abundances of some elements with respect to the average enrichment rate of the interstellar medium (blue curve in the bottom plot) and a group of stars for which the relation between the abundances of all elements is nearly constant  (purple curve in the bottom plot), this implies [$X$/Fe] $\sim0$.}   
\label{representation_curves}
\end{figure}
\begin{figure*}
\centering
\includegraphics[width=80mm]{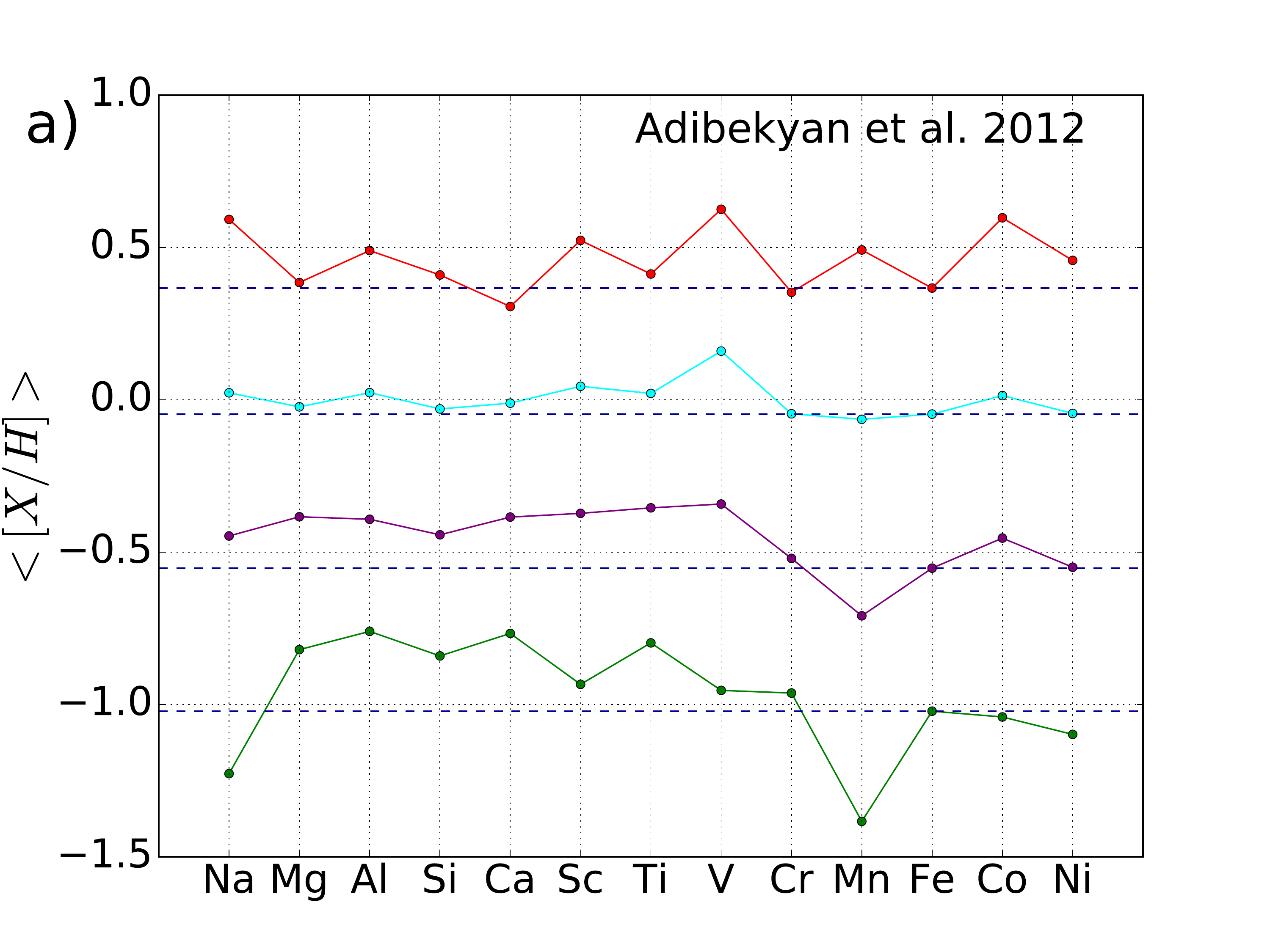}
\vspace{-0.2cm}
\includegraphics[width=80mm]{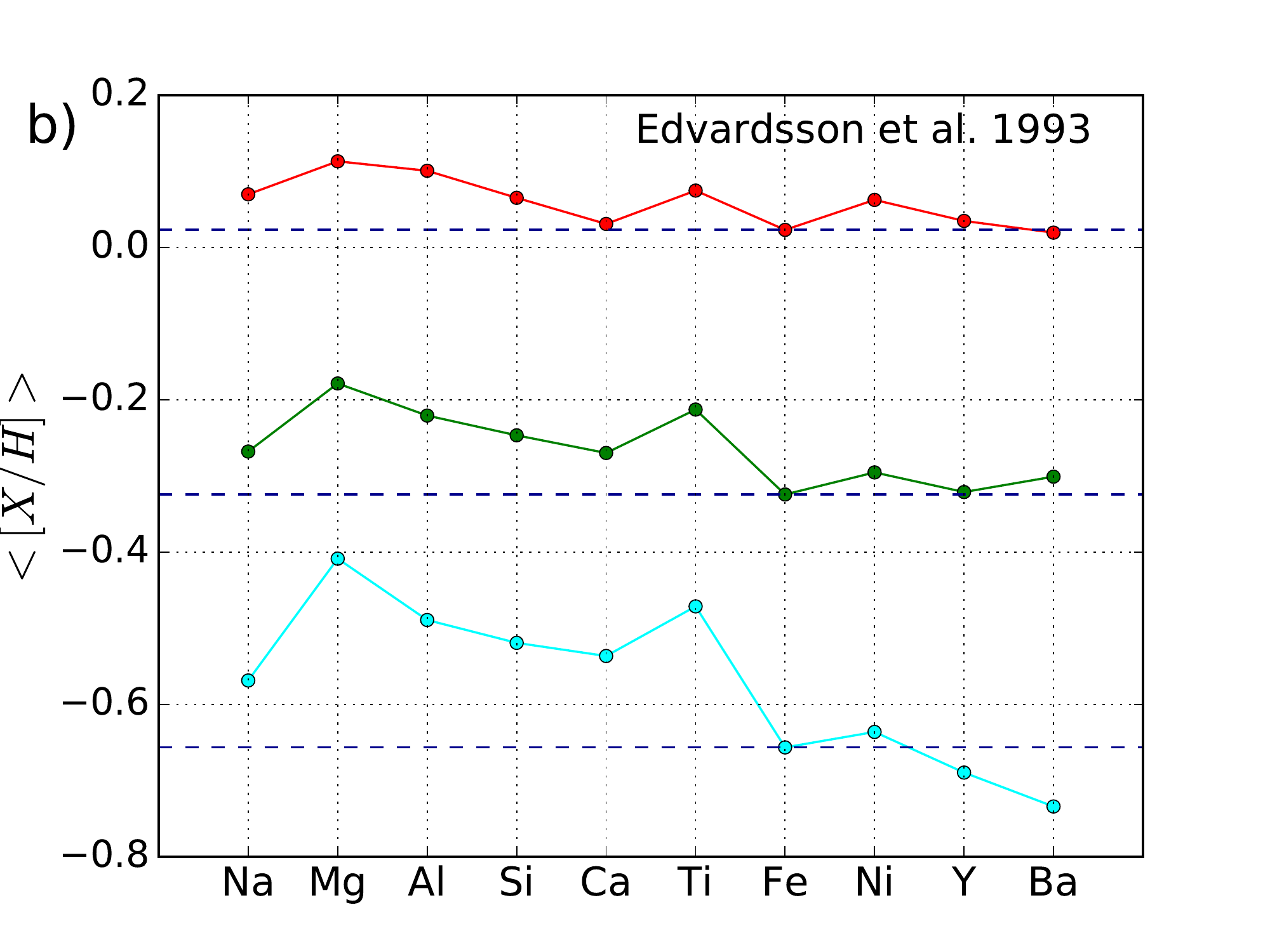}
\vspace{-0.15cm}
\includegraphics[width=80mm]{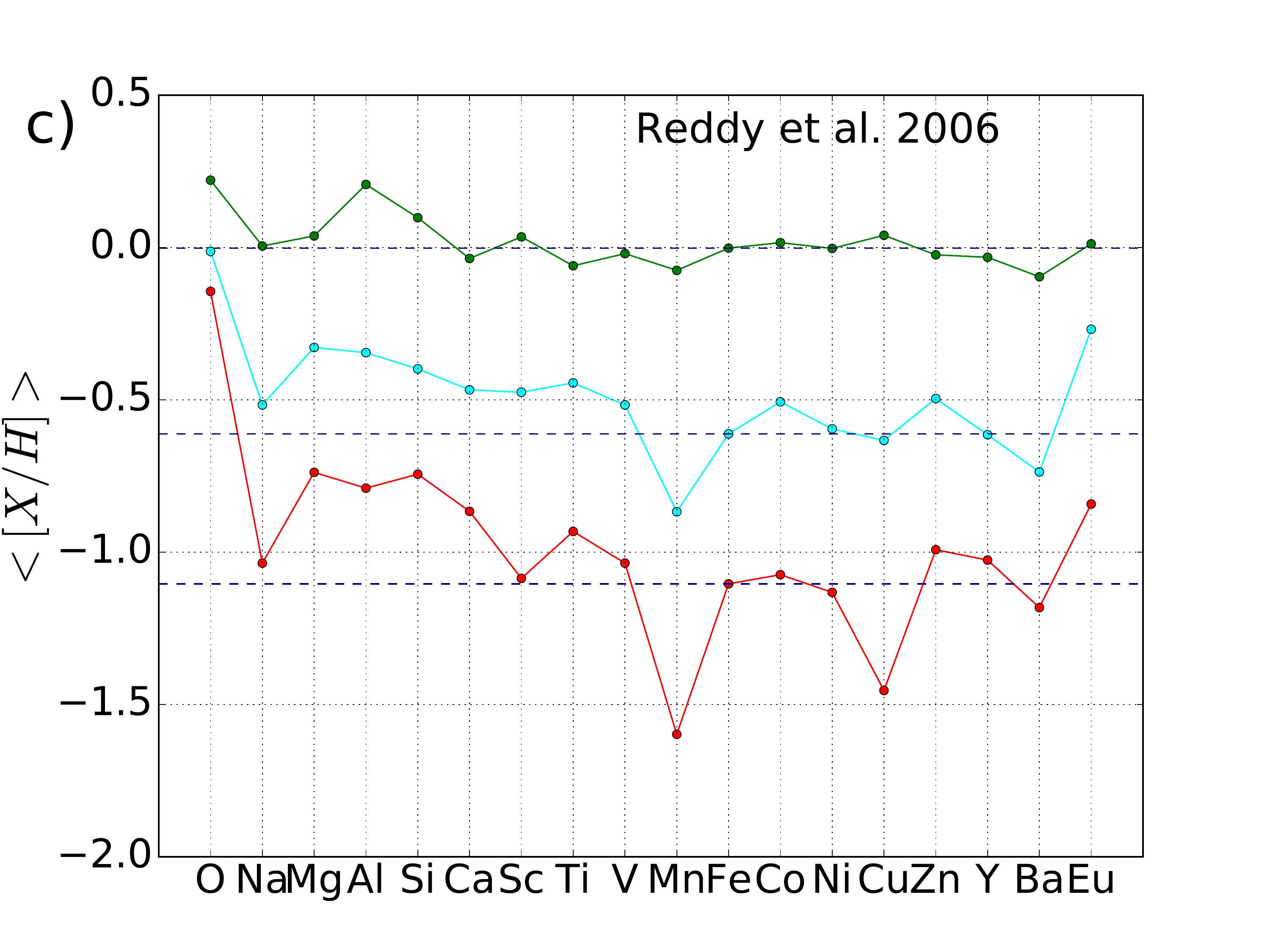}
\vspace{-0.15cm}
\includegraphics[width=80mm]{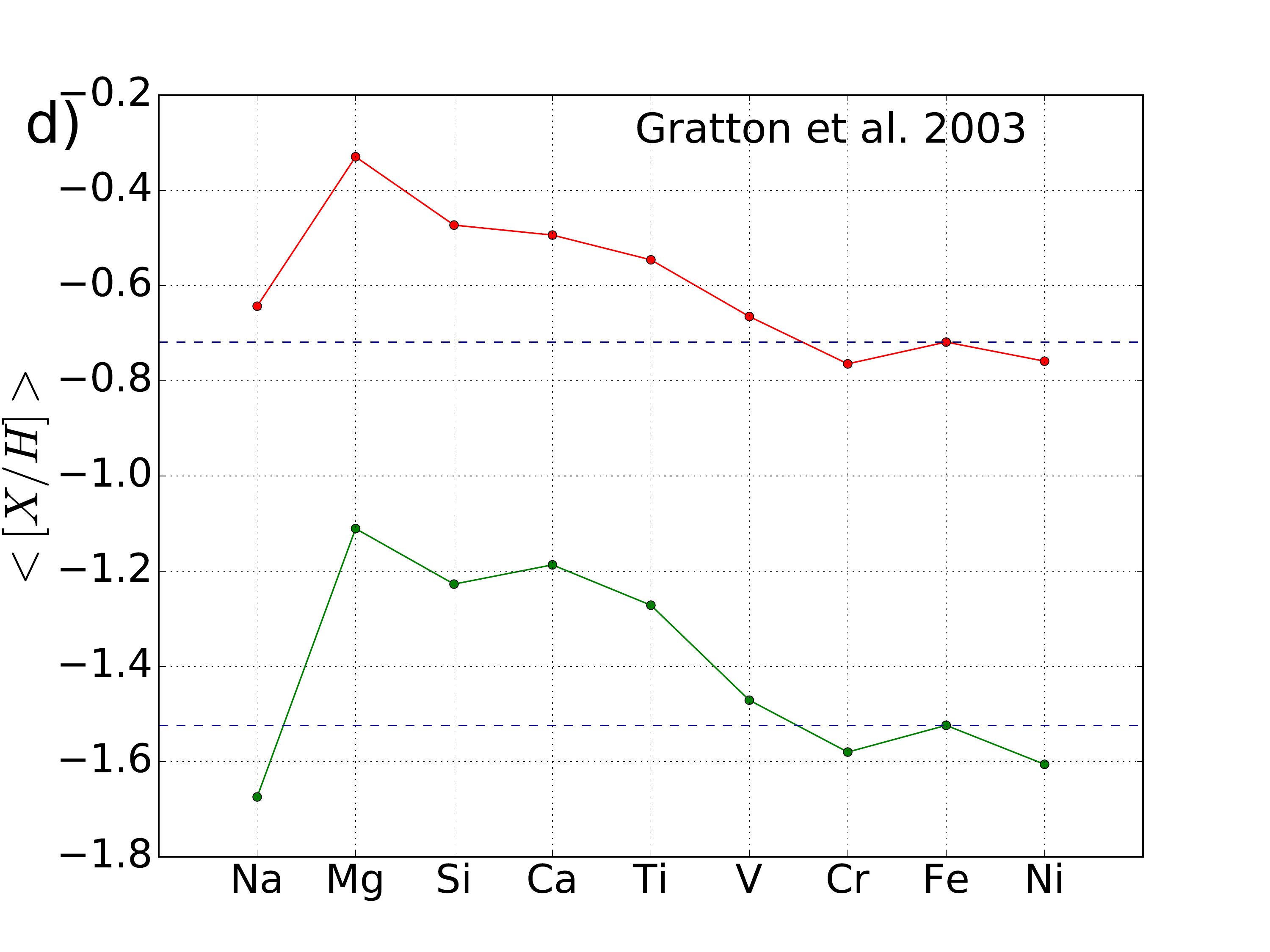}
\vspace{-0.15cm}
\includegraphics[width=80mm]{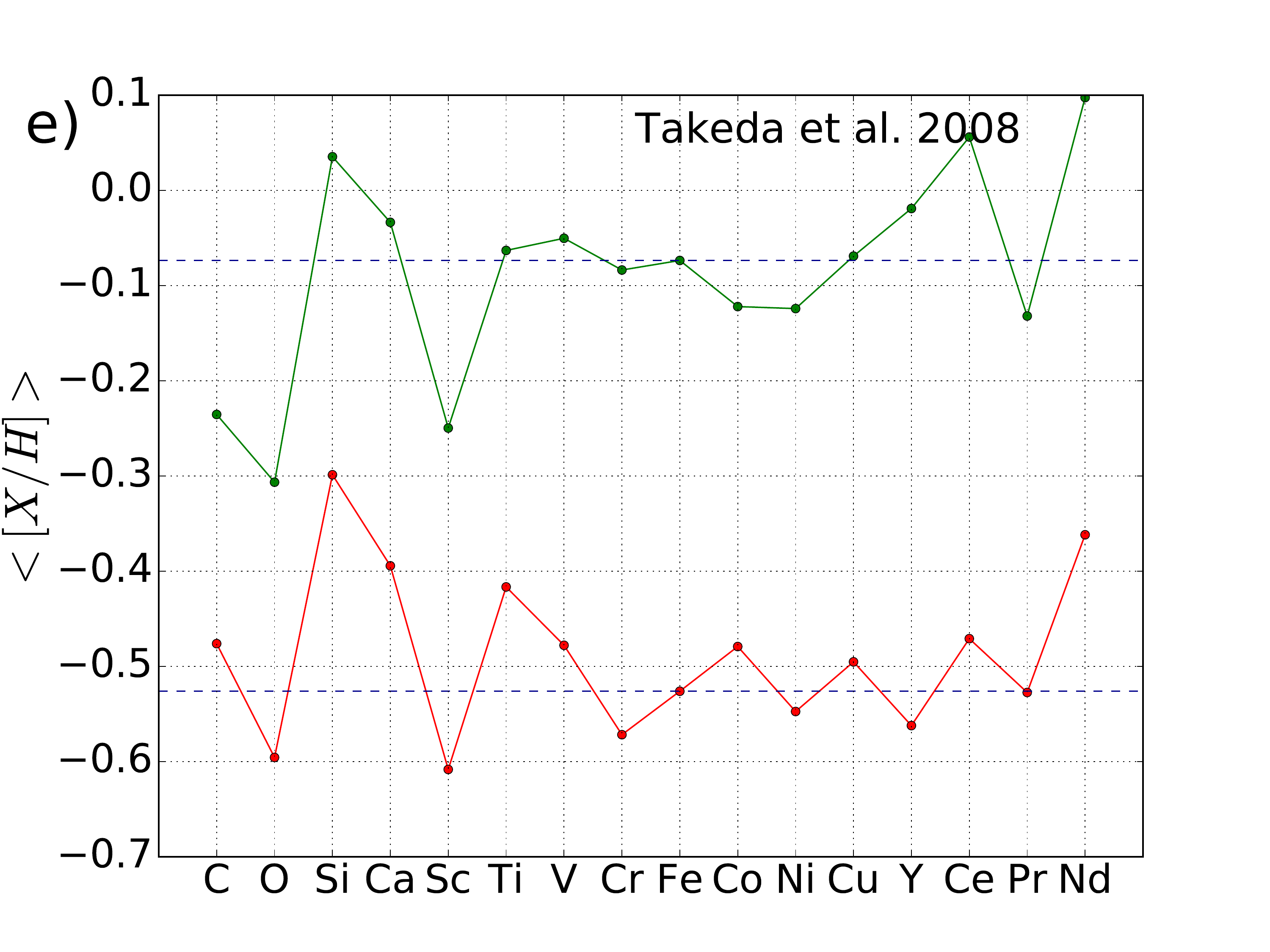}
\vspace{-0.15cm}
\includegraphics[width=80mm]{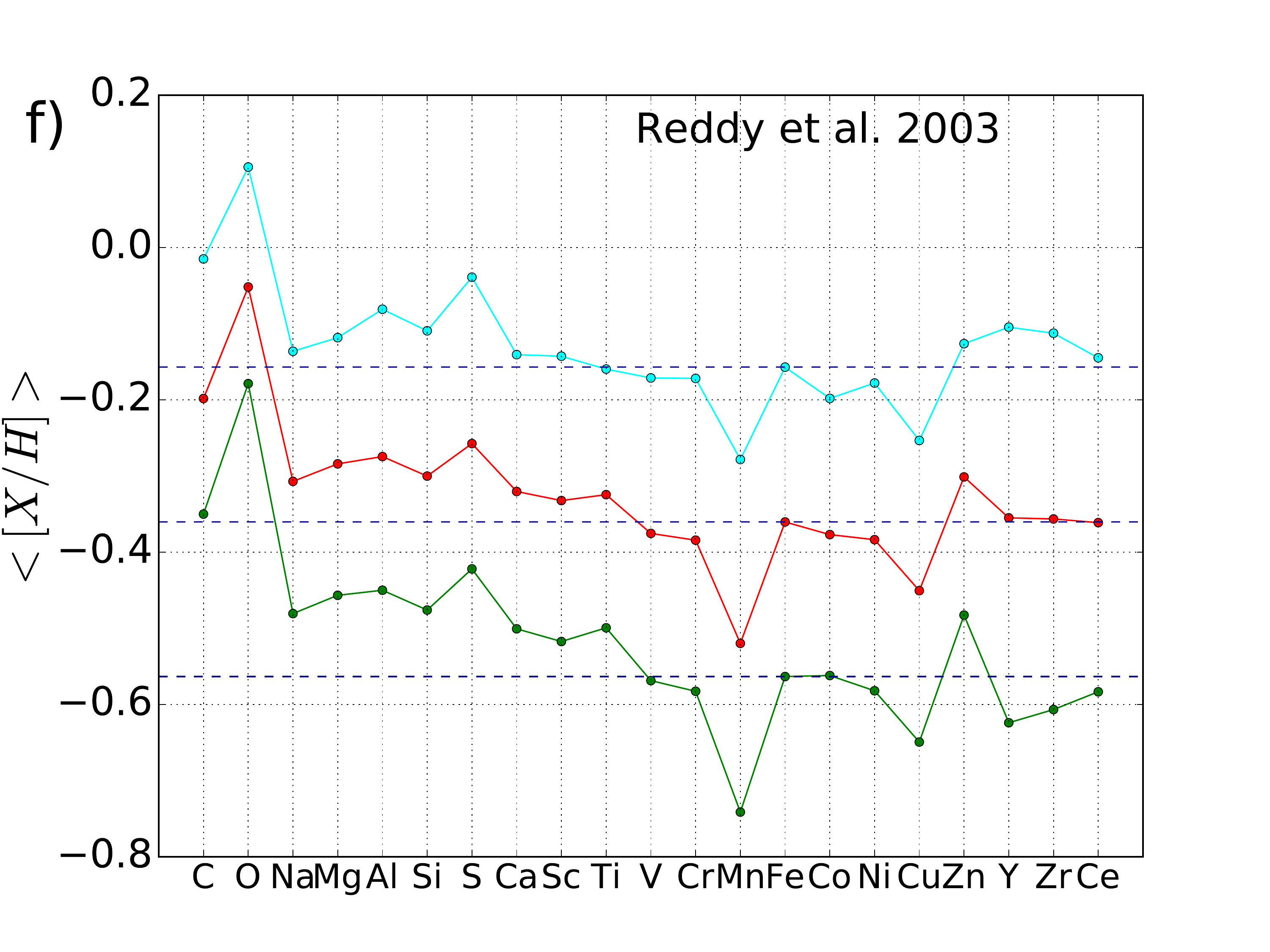}
\vspace{-0.15cm}
\includegraphics[width=80mm]{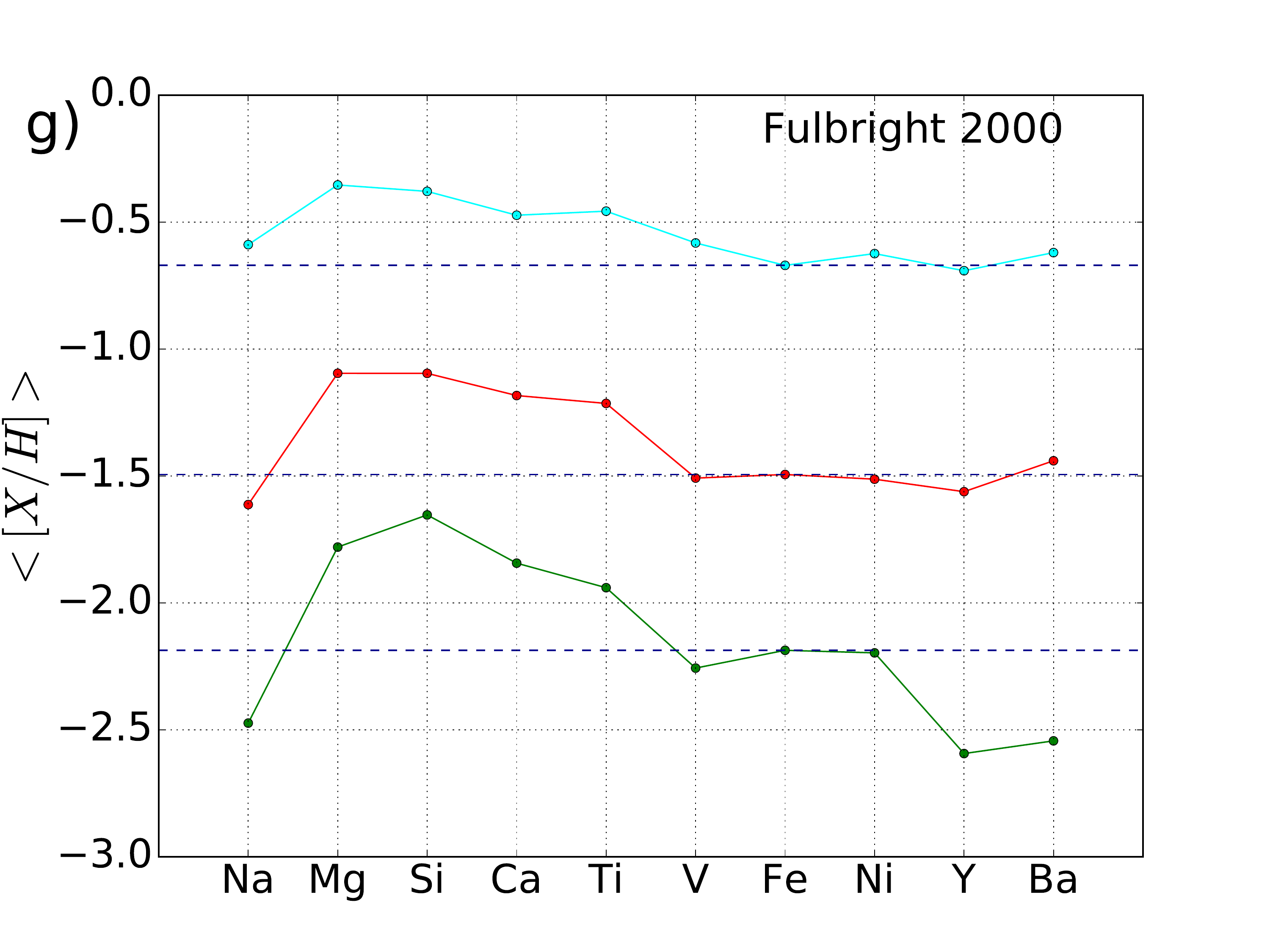}
\includegraphics[width=80mm]{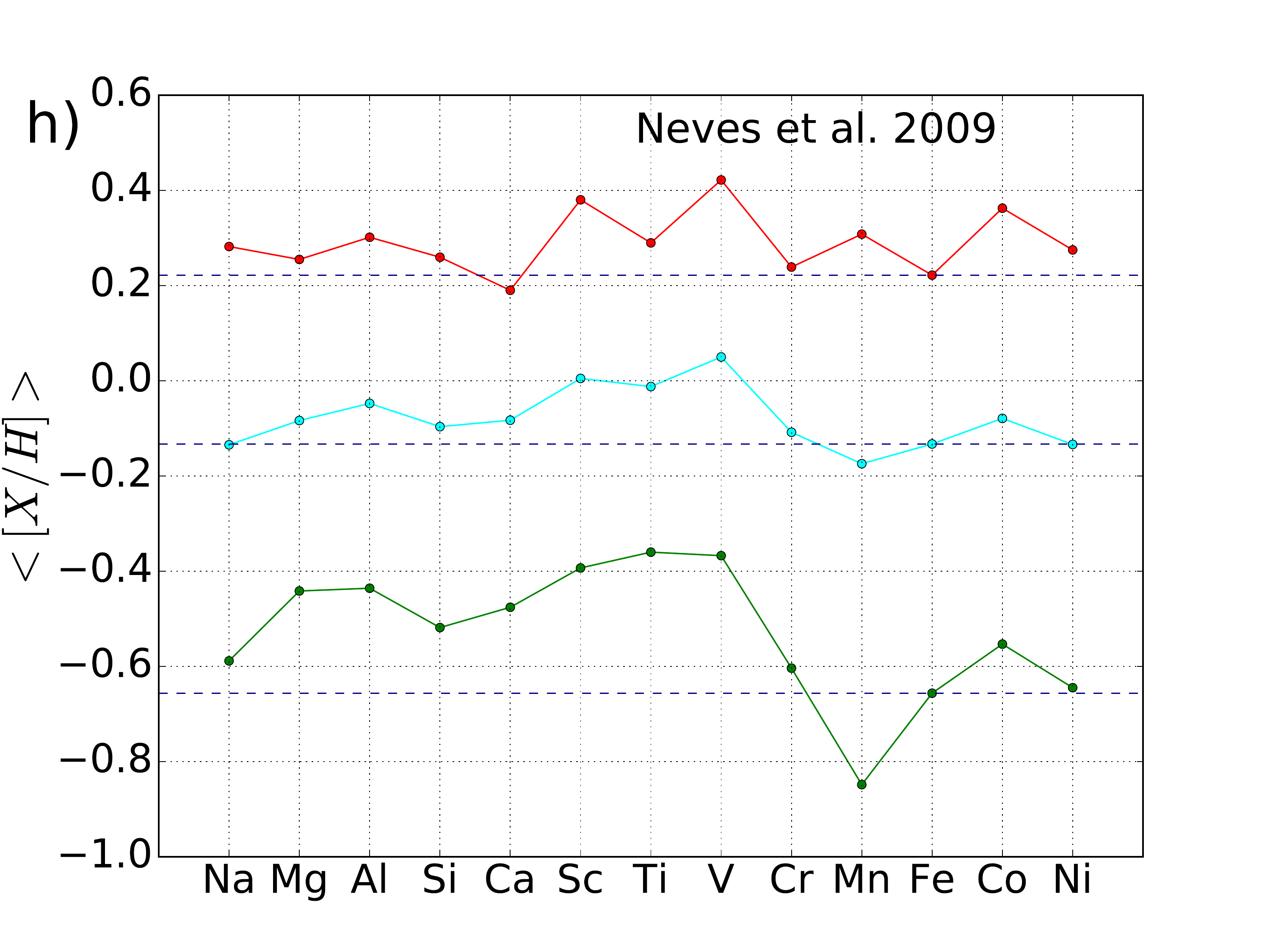}
\caption{Average abundance curves for the groups separated by hierarchical clustering for the 8 samples studied in this work. The dark blue dashed lines mark $\langle[\rm{Fe/H}]\rangle$ for each group. The colours of the average abundance curves for the groups follow the same colours used in the dendrogram that represents the hierarchical clustering analysis (see Figure \ref{tree_edv93} as example to \citet{edv93} sample).}
\vspace{-25.1pt}
\label{Abundancia_media}
\end{figure*}

\indent We seek to find clustering of stars with similar abundance patterns. To do this, we made cuts in the dendrograms in order to divide them into chemically similar groups. The cut should be done following some criteria that quantify the relevance of the groups obtained. The cut is performed after we establish a threshold value $\theta$ for the y-axis, which represents the distance between groups in the dendrogram, and from where we can infer the level of similarity of the objects. This value $\theta$ can be described as a threshold value in which all distances between different groups are greater than $\theta $ and all distances of members (subgroups and objects) within a same group are smaller than or equal to $\theta$. It is the equivalent of making a cut with a horizontal line in the dendrogram for $\delta=\theta$, where $\delta$ represents the height on the y-axis (see Figure \ref{tree_edv93}). We use a cut-off criteria to a value of $\theta$ whose number of groups remains constant over a greater range of abundance distance \citep[see Figure \ref{decaimento_edv3} and][]{placco2007}. Our dendrograms are coloured in order to distinguish each of the groups. The cut is meant to separate the first-level structural division of the stars in the abundance space. These groups can be used to order the stars in the direction defined by the chemical enrichment flow (see Section \ref{stars_peculiar_chemistry}).

\indent We define a chemical group as one these first-level divisions found using hierarchical clustering. Hereafter, every time the word group is used in the discussion, it explicitly means a group of stars that is put together in one of these first-level divisions. As the main goal of this work is to find chemically peculiar and extreme stars, we further divided the groups in subgroups, and analysed the behaviour of the average abundance of each pattern in them. After the division into subgroups, some of these were not considered in our analysis on account of the very small number of constituent stars. For the remaining subgroups we study their average abundance pattern. As with the word group, subgroup is hereafter used with an exact definition. It always means a subdivision of a previously found group.

\indent As will be shown below, we represent the abundance patterns of the groups and subgroups found with the technique of hierarchical clustering through chemical abundance curves. Each curve represents the average abundance from a group (or a subgroup) of stars with similar chemistry. In Figure \ref{representation_curves} we have two plots that are schematic representations of what we could expect when analysing the chemical abundance space. We have in these two plots chemical abundance curves for different types of groups. The first plot presents 4 groups of stars born from an interstellar medium which was enriched by a constant average rate for all elements. In this case the groups have proportional chemical abundance patterns. In the second plot we present, 2 groups that have proportional chemical abundance patterns (magenta and yellow curves in the bottom plot), a group of stars born from an interstellar medium with large variation for the abundances of some elements with respect to the average enrichment rate of the interstellar medium (blue curve in the bottom plot) and a group of stars for which the relation between the abundances of all elements is nearly constant  (purple curve in the bottom plot), this implies [$X$/Fe] $\sim0$.

\subsection{Chemical enrichment flow}

\indent We use the metallicity of stars as an approximate measure of time, so that stars having lower abundances are likely to be older since they formed from less enriched gas clouds.

\indent By applying the hierarchical clustering technique and our cut-off criteria, we find that the \citet{adibekyan2012} sample is made of four main groups, the \citet{gratton2003} and \citet{takeda2008} samples are made of two main groups while all other samples are made of three main groups. 

\indent In Figure \ref{Abundancia_media} we show the average abundance curve for each group found in the 8 samples. From the equations of chemical evolution  (see Section \ref{chemical_evolution}) we expected a progressive enrichment of the interstellar medium with time in which the elements had their abundances enhanced in a nearly constant rate. The comparison of the average abundance curves are very useful to display the relative enrichment rate of each element in what otherwise would require a multidimensional plot (one element in each dimension). 

\indent We classify these groups according to their average abundance. The abundance patterns for each group in Figure \ref{Abundancia_media} can be classified roughly in metal-poor, intermediate metallicity and metal-rich, through chemical abundance curves. The terms metal-rich, metal-poor and intermediate metallicity are used in a relative sense between groups, since the metallicity coverage of each sample differs. This result would be, by itself, trivial except for the fact that the chemical abundance patterns of the groups, in general, are somewhat proportional. They are not completely different chemical groups, but similar in the sense that what segregates them is an approximately linear increase in the abundances of all the elements together. In short, the average enrichment rate of the interstellar medium is nearly constant for all the elements.

\indent The enrichment patterns observed in Figure \ref{Abundancia_media} has analogous behaviour to the diagonal (or principal axis) of a multivariate abundance space (see Figure \ref{esquema}). We show below that this behaviour is easily understood from a principal component analysis. Due to the highly structured nature of the $C$-space, all agglomerative methods will initially segregate groups along this main enrichment flow. 

\subsubsection{Relative enrichment rate among the elements}

\indent We observe variations in the enrichment rate within the same group, whose average abundance relative to iron varies from element to element, and is due to different processes of nucleosynthesis. In all samples we observed an overabundance of the $\alpha$-elements ([$\alpha$/Fe] $>0.0$) to all groups, but there is a decrease of [$\alpha$/Fe] in relation to the increase of metallicity. This behaviour is due to the delay in the explosions of SNe Ia with respect to SNe II.

\indent The stellar abundances in \citet[see Figure \ref{Abundancia_media}a]{adibekyan2012} and also \citet[see Figure \ref{Abundancia_media}h]{neves2009} have [Mn/Fe] subsolar for [Fe/H] $<0.0$, while for [Fe/H] $>0.0$ it becomes slightly supersolar. A similar behaviour is observed in \cite{n00}, \cite{prochaska2000} and \cite{daSilva2012}. This increase of [Mn/Fe] with metallicity is explained in \cite{kobayashi2006}, through models of nucleosynthetic yields, which shows that the ratio [Mn/Fe] is increasing rapidly because a higher amount of Mn is produced by SNe Ia than Fe.


\indent Based on the plots of average abundance of the groups (see Figures \ref{Abundancia_media}b, e, f and g), we verify a significant increase in the abundances of s-process elements (Ba, Y, Zr and Ce), with the increase of metallicity ($\frac{d\langle[\rm{s/Fe}]\rangle}{d\langle[\rm{Fe/H}]\rangle}>0$) in the samples of \cite{edv93}, \cite{takeda2008}, \cite{reddy03} and \cite{fulbright2000}. Similar behaviour is observed by \cite{daSilva2012}. This significant and later enrichment occurs probably because these elements are mainly produced by intermediate mass stars during the AGB phase.

\indent We observed a similar behaviour between \citet[see Figure \ref{Abundancia_media}g]{fulbright2000} and \citet[see Figure \ref{Abundancia_media}d]{gratton2003} samples: [Na/Fe] is subsolar for $[{\rm Fe/H}]<-1.0$, whereas for $[{\rm Fe/H}]>-1.0$ it becomes supersolar. A possible increase of [Na/Fe] for high metallicity is found in \cite{daSilva2012}. To higher metallicity, [Na/Fe] becomes constant ([Na/Fe] $\sim0.0$ to [Fe/H] $>-1.0$). According to \cite{kobayashi2006}, this behaviour occur due to the dependence in metallicity of the synthesis of Na.

\indent The abundances of the sample of \cite{takeda2008} showed a distinct behaviour compared to the others. \cite{takeda2008} analysed giant stars, and in this range of stellar masses the CNO cycle, together with the convection, induces the consumption of atmospheric C and O, and because of this we observe a deficit of O and C in the patterns of the groups of this sample.

\indent The abundance of the iron peak elements (Cr, Ni, Co, Cu, Zn, Mn, V and Fe) follow closely $\langle[$\rm{Fe/H}$]\rangle$ with small variations. They also had a late enrichment, since they are synthesized at SNe Ia sites.

\subsection{Peculiar subgroups}

%

\begin{figure}
\centering
\includegraphics[width=88mm, trim=0.5cm 3.5cm 0cm 4.5cm,clip=True]{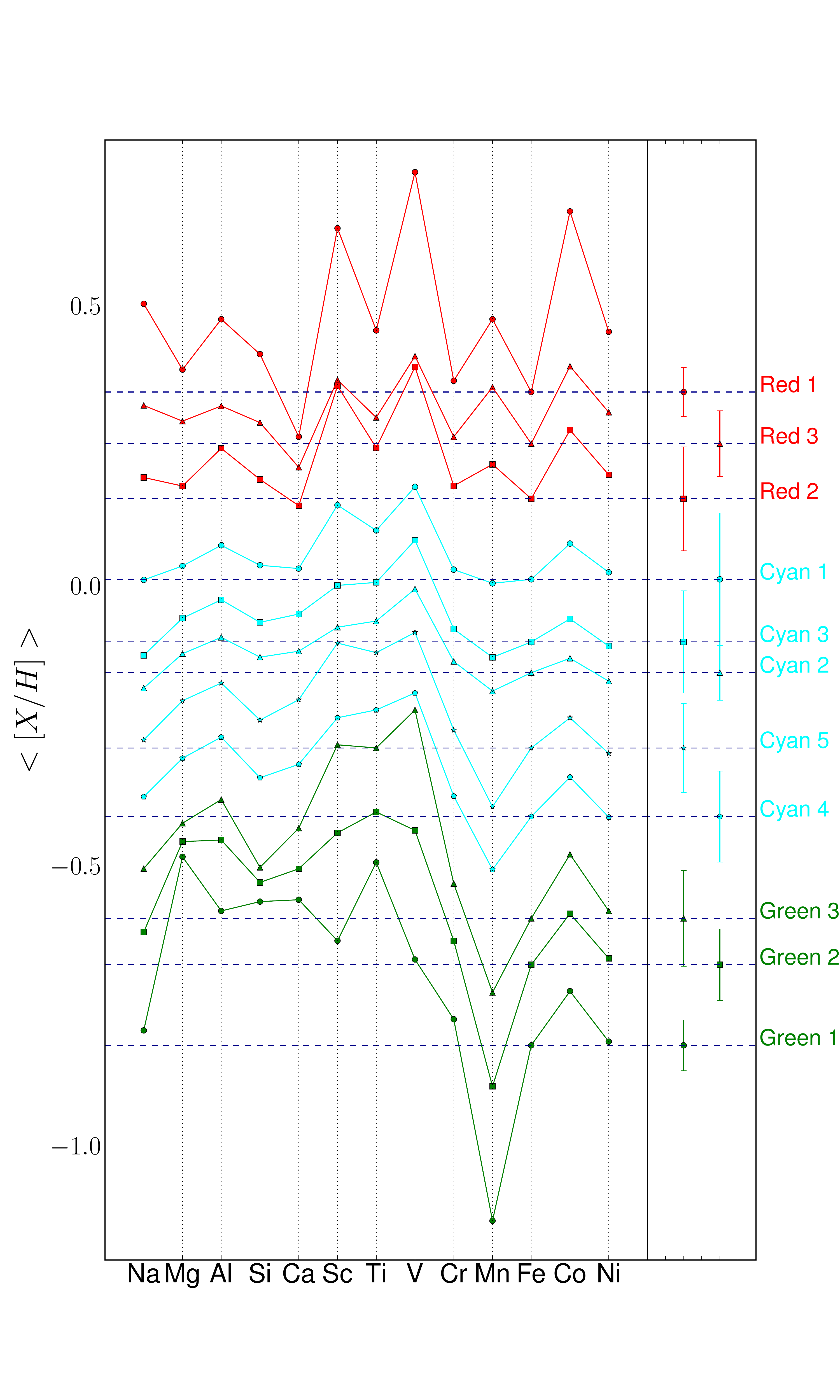}
\caption{Average abundance curves for the 11 subgroups selected from the hierarchical clustering of the \citet{neves2009} sample. The dark blue dashed lines determine the average abundance values of $\langle[$\rm{Fe/H}$]\rangle$ for each subgroup. At the right side of the plot we give the average abundance dispersion ($\sigma$) for each subgroup and an identification label according to the colour and position of the subgroup in the dendrogram that represents the hierarchical clustering analysis (see Figure \ref{tree_edv93} as example to \citet{edv93} sample).}    
\label{Abundancia_media_subgrupos_neves2009}
\end{figure}

\begin{figure}
\centering
\includegraphics[width=88mm, trim=0.5cm 3.5cm 0cm 4.5cm,clip=True]{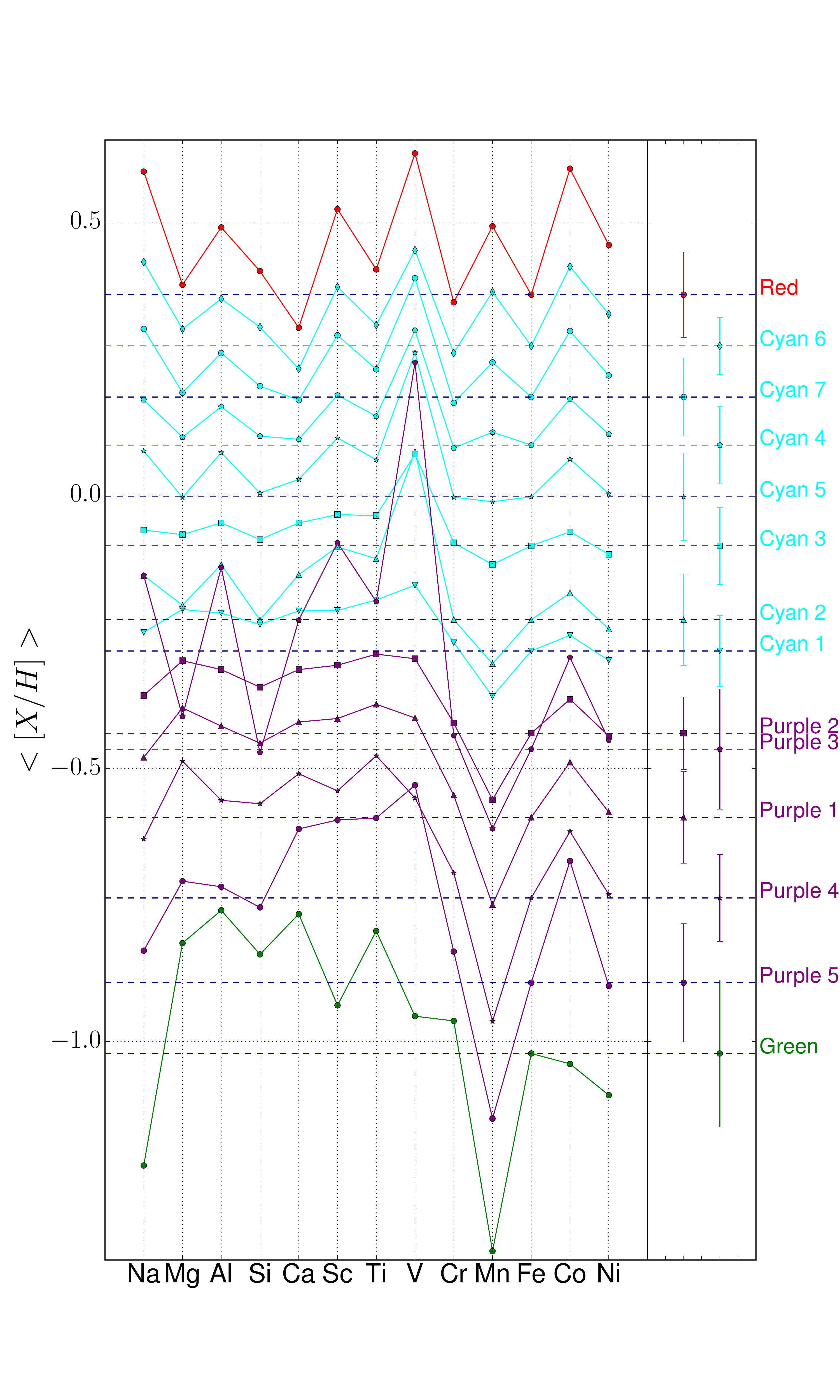}
\caption{Average abundance curves for the 14 subgroups selected from the hierarchical clustering of the \citet{adibekyan2012} sample. The dark blue dashed lines determine the average abundance values of $\langle[$\rm{Fe/H}$]\rangle$ for each subgroup. At the right side of the plot we give the average abundance dispersion ($\sigma$) for each subgroup and an identification label according to the colour and position of the subgroup in the dendrogram that represents the hierarchical clustering analysis (see Figure \ref{tree_edv93} as example to \citet{edv93} sample).}
\label{Abundancia_media_subgrupos_adibekyan2012}
\end{figure}

\begin{figure}
\centering
\includegraphics[width=88mm, trim=0.5cm 2.8cm 0cm 3.8cm,clip=True]{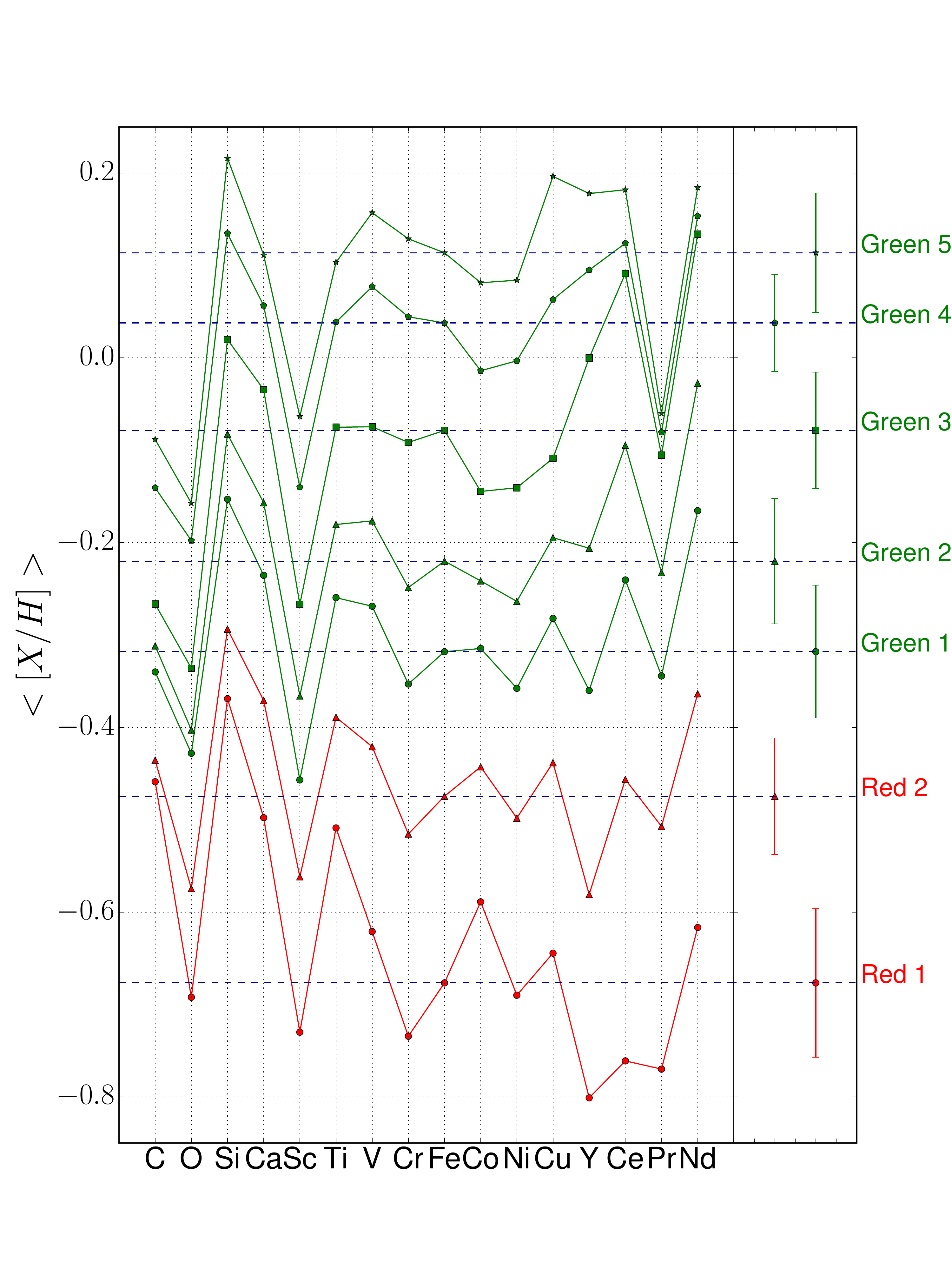}
\caption{Average abundance curves for the 7 subgroups selected from the hierarchical clustering of the \citet{takeda2008} sample. The dark blue dashed lines determine the average abundance values of $\langle[$\rm{Fe/H}$]\rangle$ for each subgroup. At the right side of the plot we give the average abundance dispersion ($\sigma$) for each subgroup and an identification label according to the colour and position of the subgroup in the dendrogram that represents the hierarchical clustering analysis (see Figure \ref{tree_edv93} as example to \citet{edv93} sample).}   
\label{Abundancia_media_subgrupos_takeda2008}
\end{figure}

\begin{figure}
\centering
\includegraphics[width=88mm, trim=0.5cm 2.8cm 0cm 3.8cm,clip=True]{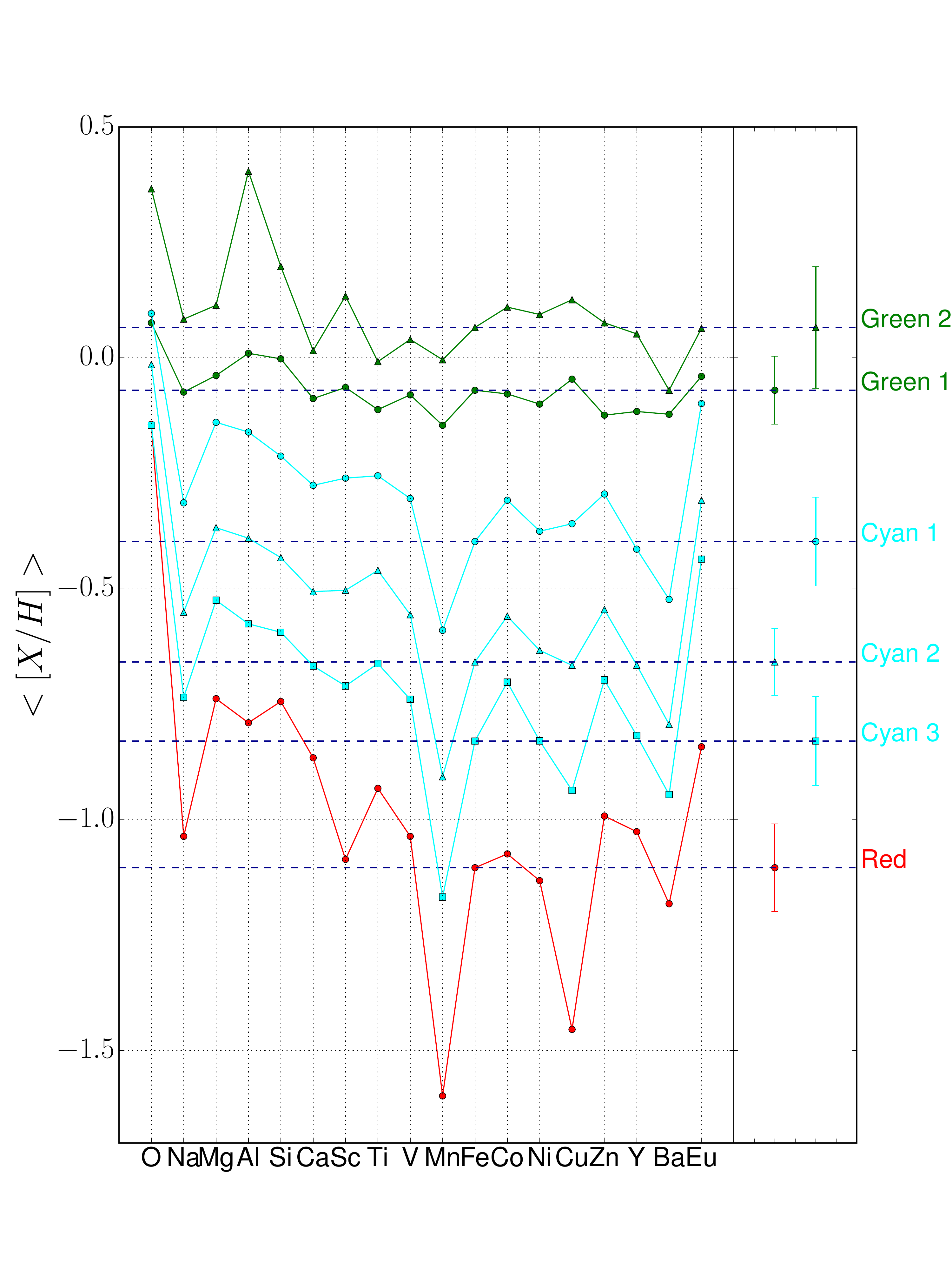}
\caption{Average abundance curves for the 6 subgroups selected from the hierarchical clustering of the \citet{reddy2006} sample. The dark blue dashed lines determine the average abundance values of $\langle[$\rm{Fe/H}$]\rangle$ for each subgroup. At the right side of the plot we give the average abundance dispersion ($\sigma$) for each subgroup and an identification label according to the colour and position of the subgroup in the dendrogram that represents the hierarchical clustering analysis (see Figure \ref{tree_edv93} as example to \citet{edv93} sample).}    
\label{Abundancia_media_subgrupos_reddy2006}
\end{figure}

\begin{figure}
\centering
\includegraphics[width=88mm, trim=0.5cm 2.8cm 0cm 3.8cm,clip=True]{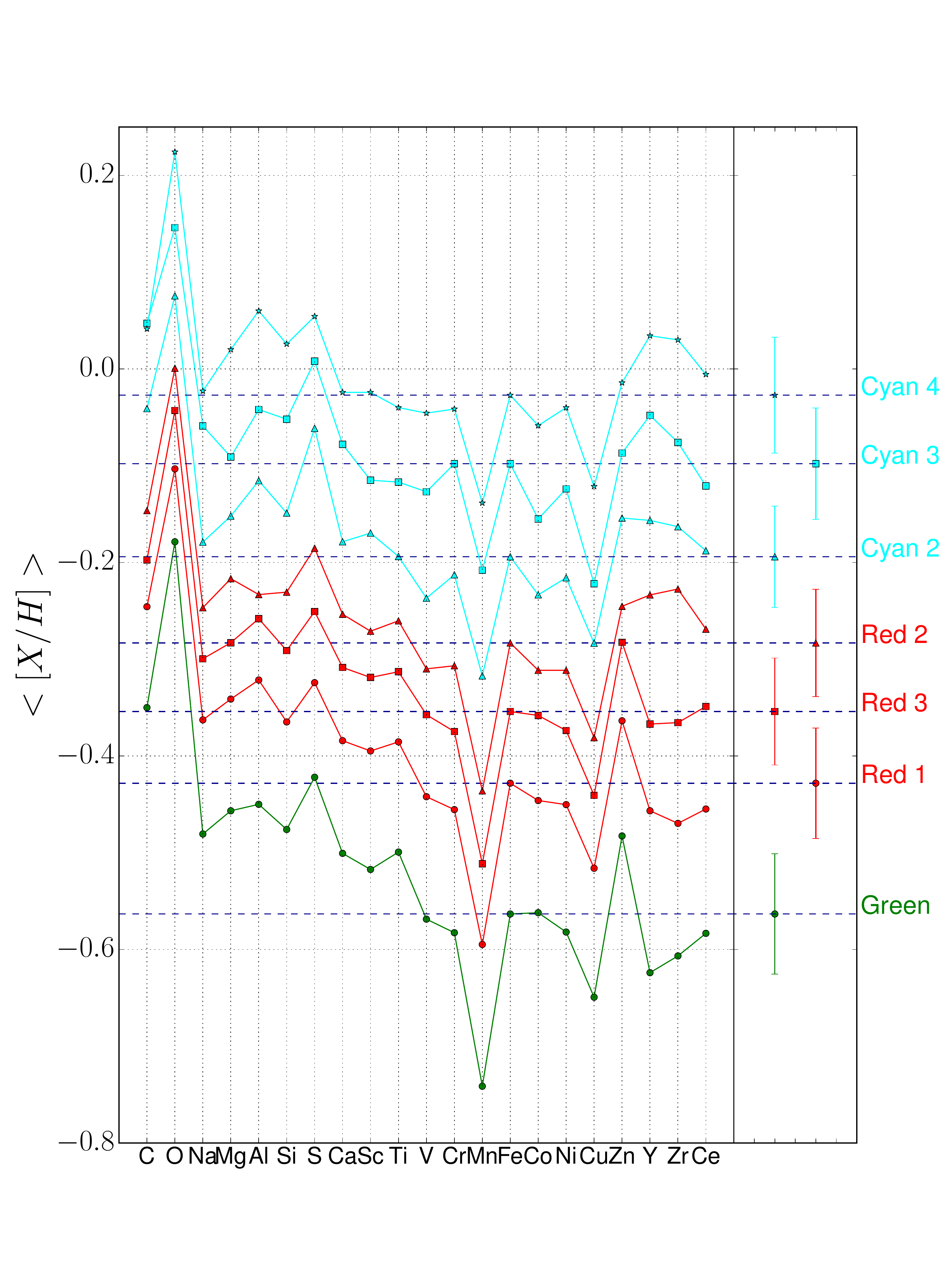}
\caption{Average abundance curves for the 7 subgroups selected from the hierarchical clustering of the \citet{reddy03} sample. The dark blue dashed lines determine the average abundance values of $\langle[$\rm{Fe/H}$]\rangle$ for each subgroup. At the right side of the plot we give the average abundance dispersion ($\sigma$) for each subgroup and an identification label according to the colour and position of the subgroup in the dendrogram that represents the hierarchical clustering analysis (see Figure \ref{tree_edv93} as example to \citet{edv93} sample).}   
\label{Abundancia_media_subgrupos_reddy03}
\end{figure}

\indent Stars that are chemically more distinct are highlighted as subgroups along the chemical enrichment flow. Because this, we decided to divide the groups of stars into subgroups with a smaller number of stars.

\indent The chemical abundance curves of the subgroups of stars in each sample are shown in Figures \ref{Abundancia_media_subgrupos_neves2009} to \ref{Abundancia_media_subgrupos_reddy03}. We label the subgroups according to the colour and position of the subgroup in the dendrogram that represents the hierarchical clustering analysis (see Figure \ref{tree_edv93} as example to \citet{edv93} sample).

\indent Large variations in abundance in the interstellar medium over time give rise to extreme subgroups with different enrichment patterns compared to the average pattern. We say that these subgroups have \textquotedblleft peculiar\textquotedblright \ chemical enrichment. These subgroups with peculiar enrichment can be quickly identified in the plots depicting the abundance curves (Figures \ref{Abundancia_media_subgrupos_neves2009} to \ref{Abundancia_media_subgrupos_reddy03}). In these peculiar subgroups, the abundance of certain elements will grow at a rate faster than expected by the main chemical enrichment flow. If that happens, the curve of a subgroup with lower average abundances might cross the curve of a subgroup with higher average abundance at that specific element that has peculiar abundances.

\indent In all samples, as was expected, the main subgroups follow the chemical enrichment flow. We found that the samples of \citet{edv93}, \citet{fulbright2000} and \citet{gratton2003} do not have any element with peculiar enrichment. The chemical pattern of the subgroups identified in this sample basically show the decrease in the average value of [$\alpha$/Fe] due to the increase of [Fe/H]. The small number of stars of these samples (less than 190 stars) leads us to believe that this is the result of a small sample, statistically unrepresentative of rarer groups. Because this, we decide do not present in this paper the chemical abundance curves of the subgroups for these samples.

\indent In the \cite{neves2009} sample we found that the chemical pattern of some subgroups that almost intersect others. In this sample, 3 of 11 subgroups (see Figure \ref{Abundancia_media_subgrupos_neves2009}) have peculiar chemical patterns. There are two subgroups overabundant in Sc and V: one with average metallicity $\langle$[Fe/H]$\rangle$ = $-$0.59 dex (\textquotedblleft Green 3\textquotedblright \ subgroup, 9 stars, average abundance dispersion $\sigma$ = 0.085 dex) and one with average metallicity $\langle$[Fe/H]$\rangle$ = 0.159 dex (\textquotedblleft Red 2\textquotedblright \ subgroup, 34 stars, average abundance dispersion $\sigma$ = 0.093 dex). There is one subgroup overabundant in Mg with average metallicity $\langle$[Fe/H]$\rangle$ = $-$0.817 dex (\textquotedblleft Green 1\textquotedblright \ subgroup, 3 stars, average abundance dispersion $\sigma$ = 0.045 dex).

\indent A similar behaviour was observed in the \citet[see Figure \ref{Abundancia_media_subgrupos_adibekyan2012}]{adibekyan2012} sample. In this sample, 3 of 14 subgroups have peculiar chemical patterns. There are two subgroup overabundant in V: one with average metallicity $\langle$[Fe/H]$\rangle$ = $-$0.228 dex (\textquotedblleft Cyan 2\textquotedblright \ subgroup, 100 stars, average abundance dispersion $\sigma$ = 0.084 dex) and one with average metallicity $\langle$[Fe/H]$\rangle$ = $-$0.893 dex (\textquotedblleft Purple 5\textquotedblright \ subgroup, 8 stars, average abundance dispersion $\sigma$ = 0.108 dex). There is one subgroup overabundant in Na, Al, Ca, Sc, Ti and V with average metallicity $\langle$[Fe/H]$\rangle$ = $-$0.463 dex (\textquotedblleft Purple 3\textquotedblright \ subgroup, 12 stars, average abundance dispersion $\sigma$ = 0.11 dex).

\indent Based on the average abundance uncertainties in the \cite{adibekyan2012} and \cite{neves2009} samples, 0.03 dex and 0.02 dex for [Fe/H], respectively, and the average abundance dispersions for the groups found as peculiar to these samples, we can conclude that the peculiar groups can be considered robust groups in the stellar abundance space and not a mere effect of high values of abundance uncertainties.

\indent We observe similar trend in the \citet[see Figure \ref{Abundancia_media_subgrupos_takeda2008}]{takeda2008}, \citet[see Figure \ref{Abundancia_media_subgrupos_reddy2006}]{reddy2006} and \citet[see Figure \ref{Abundancia_media_subgrupos_reddy03}]{reddy03} samples.

\indent The samples of \cite{reddy03} and \cite{reddy2006} have a small number of objects, each about 180 stars, and probably because of this they do not present significant subgroups with peculiar abundance chemical pattern.

\subsection{Principal component analysis}

\indent In order to confront our results with other methodology, we applied the principal component analysis. The PC1 (first principal component) and PC2 (second principal component) account for more than 90$\%$ of the variance in $n$-dimensional space of stellar abundances for all samples used. This shows that the effective dimension of the abundance space is very small, which can be reasonably well-represented by 2-3 variables formed by the linear combination of the chemical abundances. Because of this, we only plot the PC1 and PC2 coordinates.

\begin{figure}
\centering
\includegraphics[width=42.7mm]{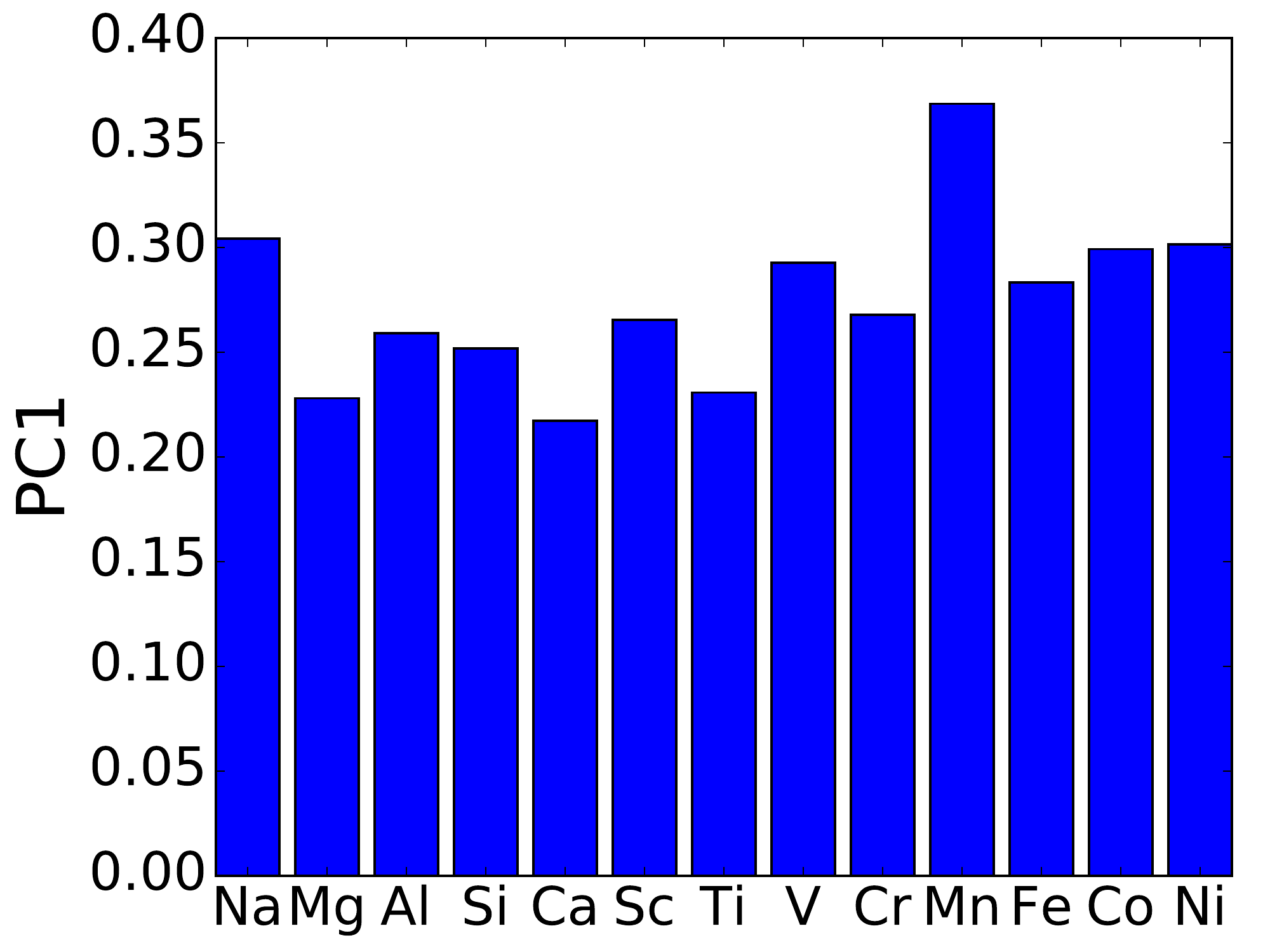}
\hspace{-3mm}
\vspace{2mm}
\includegraphics[width=42.7mm]{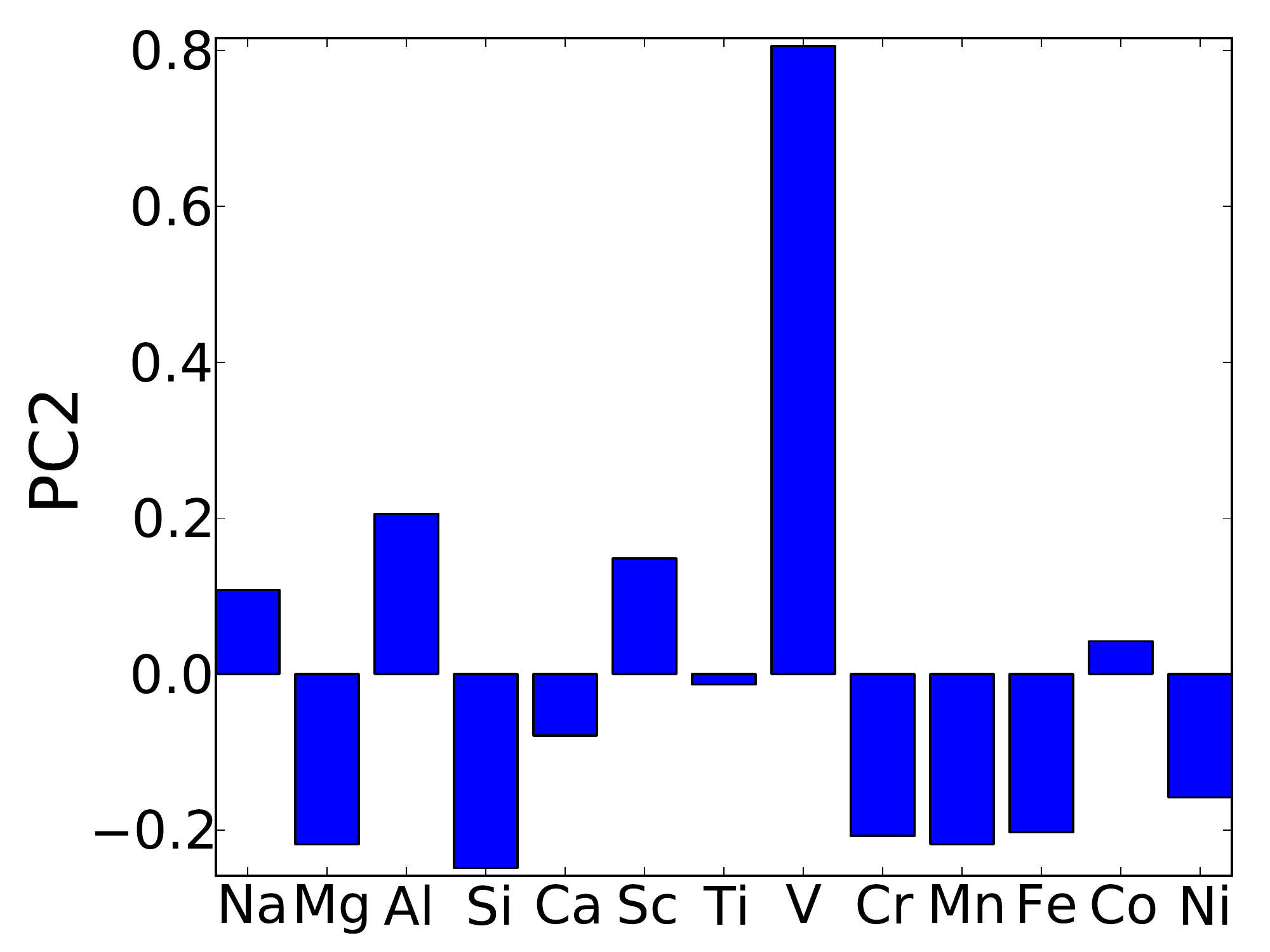}
\includegraphics[width=42.5mm]{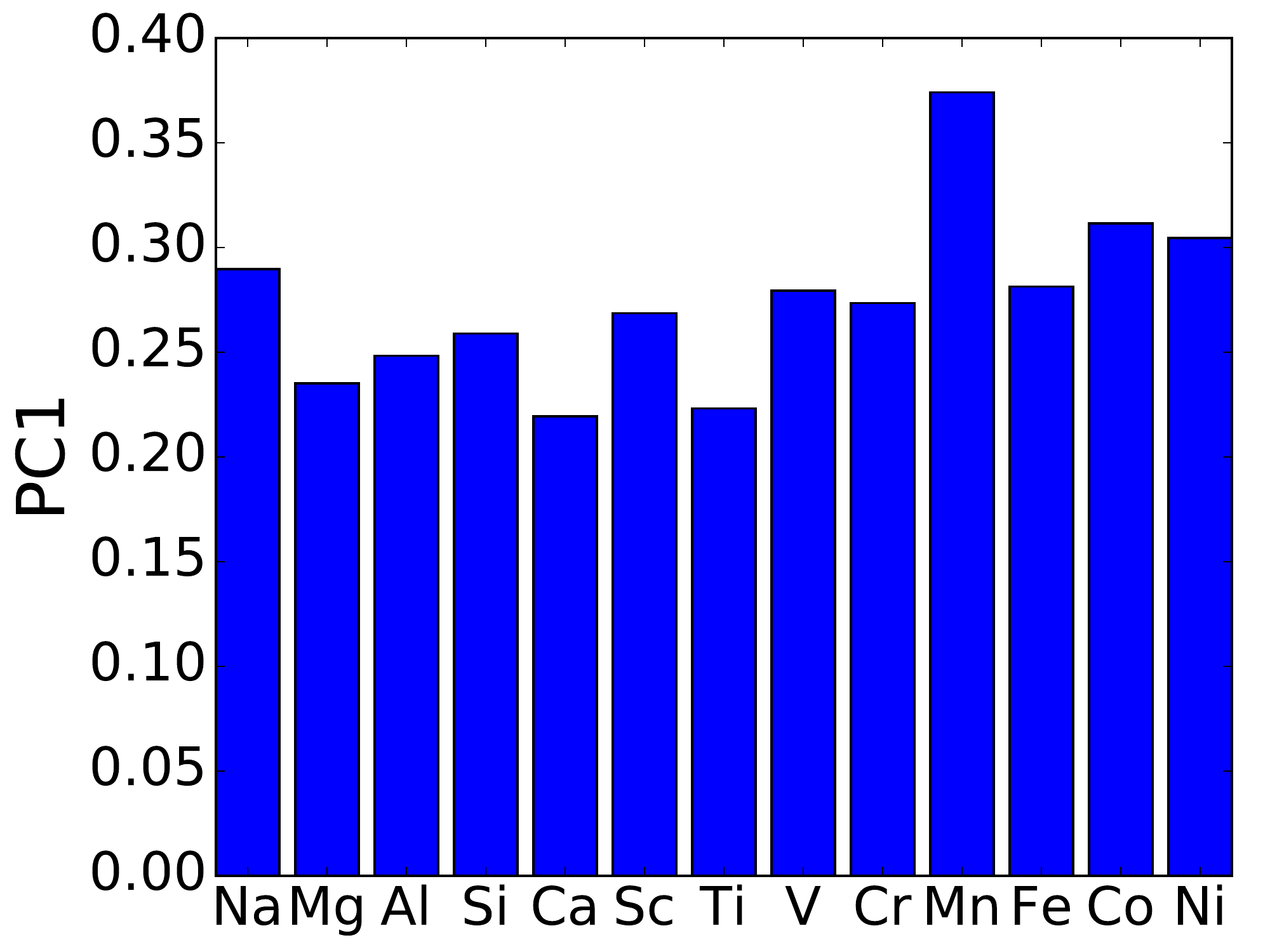}
\hspace{-2.6mm}
\includegraphics[width=42.5mm]{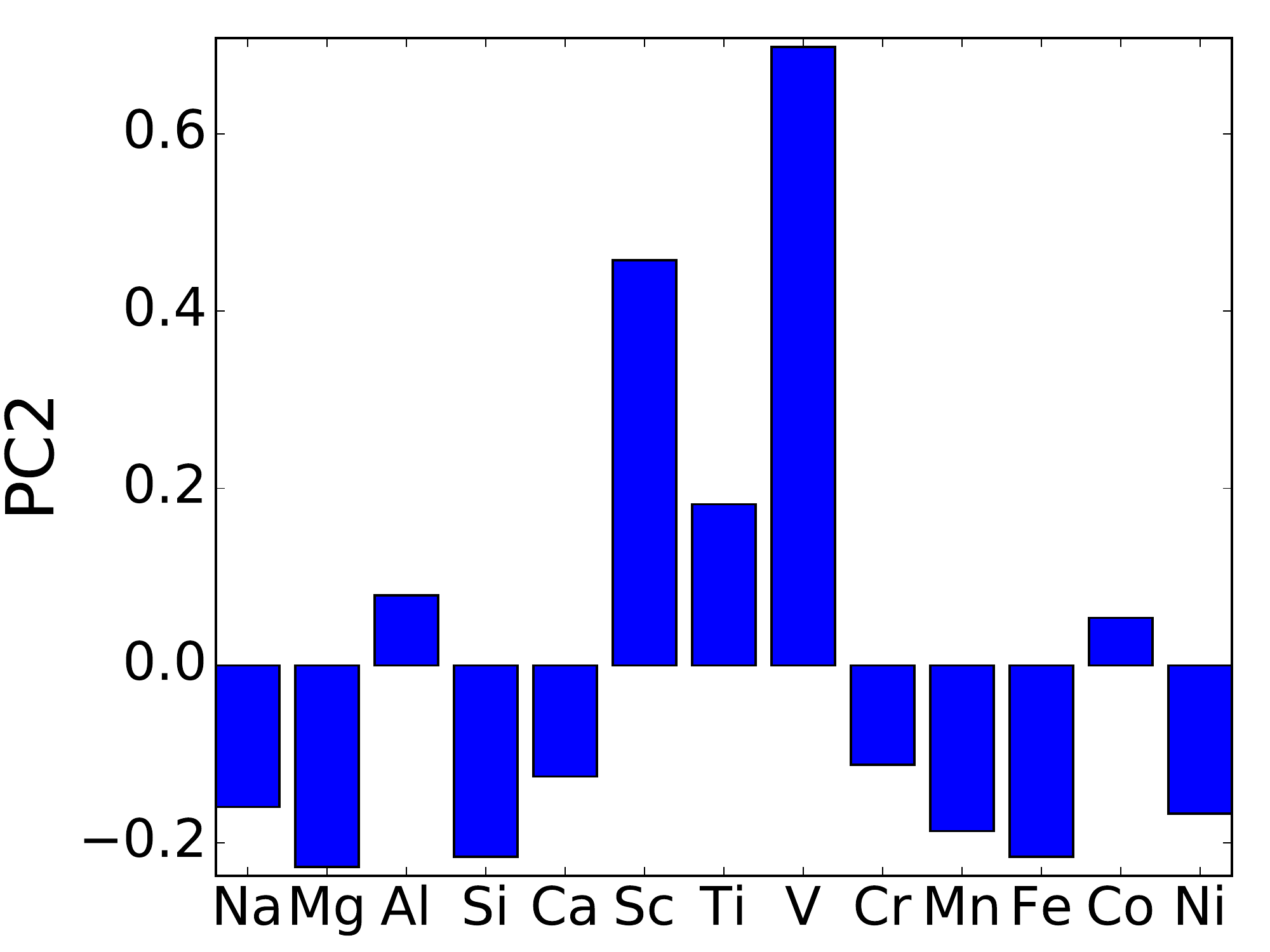}
\caption{Principal component analysis for the first (PC1) and the second (PC2) components for \citet[top panels]{adibekyan2012} and  \citet[bottom panels]{neves2009}  samples. PC1 determines the average pattern of chemical enrichment, the main flow of chemical evolution, while PC2 provides the typical deviation of this flow due to the local chemical enrichment inherited by the stars. We can see that the elements that cross in the average abundance patterns (see Figures \ref{Abundancia_media_subgrupos_adibekyan2012} and \ref{Abundancia_media_subgrupos_neves2009}) are the elements which have a greater loading in the second principal component (PC2).}    
\label{PCA1PCA2}
\end{figure}

\indent The groups found by the hierarchical clustering correspond, almost completely, to specific intervals of PC1 coefficients. As can be seen in Figure \ref{PCA1PCA2}, which corresponds to the principal component analysis for  \cite{adibekyan2012} sample, for the first and second components, respectively, PC1 determines the average chemical enrichment pattern (that is, our hierarchical groups are mostly defined by well marked intervals in PC1), the main flow of chemical evolution, while PC2 provides the typical deviation of this flow due to the local chemical enrichment inherited by stars .

\indent PC1 gives the average chemical enrichment resulting from  SNe explosions and stellar winds. PC2 represents smaller variations in the abundance space and highlights the peculiar enrichment in some elements.

\indent The principal components can be interpreted as a result of the nucleosynthesis processes. As obtained by \cite{ting2012}, who found that a dimensionality of 7 to 9 components in the $C$-space can describe 25 variables (abundances), we found that our abundances may also be represented by a lower number of components (2-3 components) than variables, in order to represent the chemical enrichment of the Galaxy.

\indent We show in Figure \ref{PCA1PCA2} that the PC1 loadings are very similar for the \cite{adibekyan2012} and \cite{neves2009} samples. The higher loadings in PC2 are seen just for those elements that mostly deviate from the typical linear increase in the average abundance patterns of the subgroups (like V, Mg, Sc and Si, see Figures \ref{Abundancia_media_subgrupos_adibekyan2012} and \ref{Abundancia_media_subgrupos_neves2009}). Similar behaviours were found in the PCAs of other samples.

\begin{figure}
\centering
\includegraphics[width=88mm,trim=0.1cm 0.2cm 0.6cm 1.4cm,clip=True]{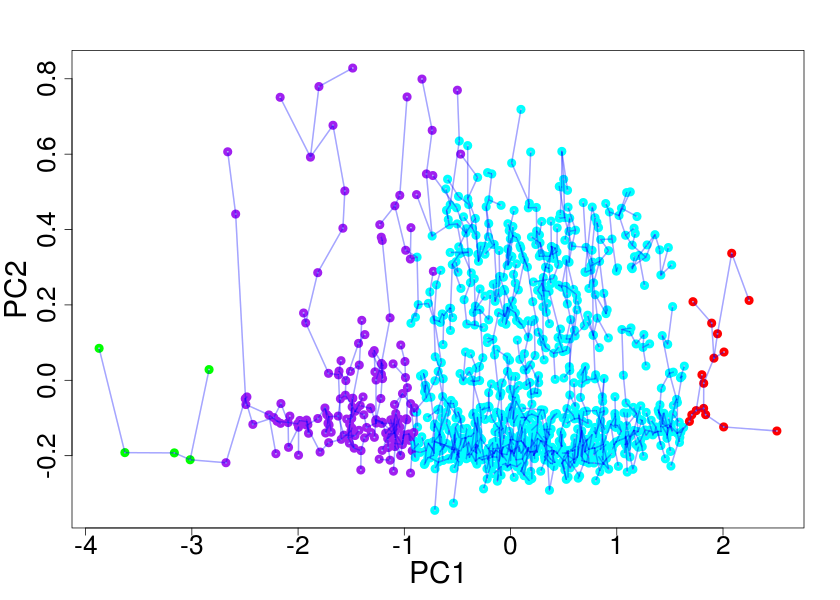}
\includegraphics[width=88mm,trim=0.1cm 0.2cm 0.6cm 1cm,clip=True]{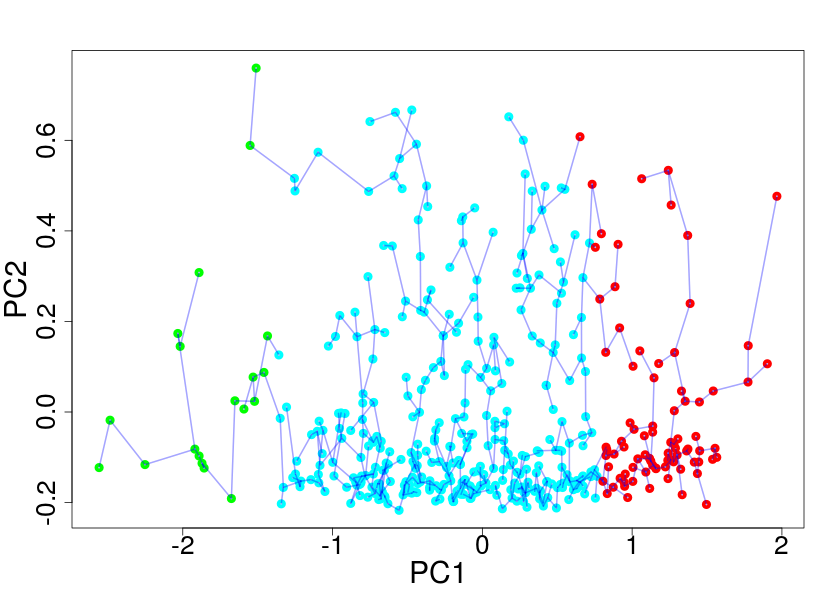}
\caption{Minimum spanning tree of the \citet{adibekyan2012} and \citet{neves2009} samples projected onto the subspace formed by the two first principal components (PC1 and PC2). The colours of the objects follow the same colours used for the hierarchical clusters found in Figure \ref{tree_edv93}. The groups are connected along a line that runs nearly parallel to the PC1 axis. PC1 marks the axis that explain most of the variance in the data. The stars are mostly aligned in low dimension elongated subspace of the $C$-space, having PC1 as its main axis. The ramifications from this flow runs parallel to the PC2. Each is a subgroup in the hierarchical clusters.}    
\label{P1P2_pca}
\end{figure}

\indent Figure \ref{P1P2_pca} shows the minimum spanning trees of the \cite{adibekyan2012} and \cite{neves2009} samples projected onto the subspace formed by the two first principal components. The colours of the objects follow the same colours used for the hierarchical clusters found in Figure \ref{tree_edv93}. The groups are connected to each other along a line that runs nearly parallel to the PC1 axis. As we have discussed before, PC1 (the first principal component) marks the axis that explain most of the variance in the data. Since the chemical enrichment in one element is generally followed by enrichment in all of the other elements, on account of the mixing of the ejecta by several stars in the interstellar medium, the stars are mostly aligned in low dimension elongated subspace of the $C$-space, having PC1 as its main axis. This is what we called \textquotedblleft main enrichment flow\textquotedblright, meaning that the chemical enrichment of the Galaxy, as seen in the C-space, processed mostly according to this path.

\indent The ramifications from this flow runs parallel to the PC2. Each is a subgroup in the hierarchical clusters. From Figure \ref{esquema}, we expect that these ramifications could evidence clusters of stars with peculiar chemistry. Indeed, these subgroups are more interesting for chemical tagging because they correspond to stars that are disproportionally over-underabundant in one or more elements.

\indent For instance, in Figure \ref{PCA1PCA2}, we see that all the PC1 loadings have the same sign (positive). Thus, a star that have coordinates (PC1,PC2) $=(-1,0)$ differs from another having coordinates (PC1,PC2) $=(+1,0)$ just by a proportional increase in all elements given by the PC1 loadings. On the other hand, the PC2 loadings have opposite signs for some elements. This means that running along the PC2 axis (for a fixed PC1 value), we will find stars that are sistematically more abundant in few elements and underabundant in others with respect to the main enrichment flow 

\indent Most of the parallel branches along the PC2 occur in the subgroups of intermediate metallicity, marked as purple and cyan in the \citet[see Figure \ref{P1P2_pca}]{adibekyan2012} sample, and as cyan in the  \citet[see Figure \ref{P1P2_pca}]{neves2009} sample. This can also be seen in the average chemical pattern of other samples where the overlapping in patterns are observed mostly in intermediate metallicity subgroups. It is likely that more branches could be found in metal-rich and poor subgroups (subgroups of extreme metallicity), if we had a larger number of stars in these regimes. 

\section{Conclusions}

\indent We used two different techniques to understand the structure of the stellar chemical abundance space. Applying the hierarchical clustering technique, we classified the stars from each sample according to their average abundances, dividing the set of stars in rich, poor and intermediate abundance groups. The PCA technique showed, to all samples, that the stellar abundances space can be reasonably well represented by only two principal components, corresponding to more than 90$\%$ of the variance in the data.

\indent The results obtained by the PCA technique corroborate with those obtained by the hierarchical clustering technique. In both, we attest the existence of a main enrichment flow in the chemical abundance space from which less prominent peculiar patterns arise.

\indent The chemical enrichment flow can be observed through graphic representation of the average abundance pattern of the groups found in the hierarchical clustering: the increase in [$X$/H] gives rise to nearly parallel patterns between the groups, so that the increase of [Fe/H] is followed by the similar growth of other abundances. This means that there is a nearly constant enrichment rate. This is equivalent to the $n$-diagonal line in the stellar abundance space. By using PCA method, we obtained a similar result. This enrichment pattern is present in all samples, and gives the bulk enrichment of the galactic interstellar medium.

\indent Ramifications out of this enrichment flow indicate peculiar chemical subgroups formed by stars with similar [Fe/H]. They show up as subgroups having intercrossing chemical patterns with subgroups of similar [Fe/H]. According to the PCA, we see that the elements that have a higher loading in the second principal component (PC2) are the same that appear with a peculiar enrichment responsible for the crossing patterns presented by hierarchical clustering technique.

\indent We obtained the same set of results through the minimum spanning tree to applied to the main components. We observed that most of the stars are along the PC1 axis, which represents the enrichment flow connecting stars of the metal-poor group to stars of the metal-rich group. Along the PC2 axis there are branches, which are the stars of subgroups with peculiar chemical enrichment.

\indent Subgroups having peculiar chemical enrichment are found mostly among subgroups of intermediate abundance. This behaviour can be easily checked through the minimum spanning tree for the samples of \cite{neves2009} and \cite{adibekyan2012}. This may be just the effect of a lower sample at the tails of the metallicity distribution. The same effect is observed in samples having small number of objects. For these, we observe just the chemical enrichment flow. The crossing patterns is observed only in the samples that have a large number of stars.

\indent We observed through our results and in the literature data, that the behaviour which we found -- intersections in the chemical abundance pattern -- results from stars overenriched in some elements. Mg, Si, Sc and V were some of the elements singled-out in the crossing abundance patterns for the \cite{neves2009} and \cite{adibekyan2012} samples. 

\indent  The peculiar enrichment of these elements for some stars may be due to production of this element from different nucleosynthetics sites. A similar behaviour to that obtained in our analysis for the Mg and Si is also found in \cite{chen2000}. \cite{daSilva2012} associate the super enrichment of the elements Mg and Si to different nucleosynthetic sites (SNe Ia and SNe II). According to the work of \cite{fenner03}, the magnesium isotopes are originated from different sites. \cite{bensby03} found that the behaviour of the Si abundance in the thick disk denotes that this element is produced by both SNe Ia and SNe II. \cite{n00} found a great enrichment of Sc in metal-poor stars, and concluded that Sc, despite being considered an iron peak element, also behaves as an element of $\alpha$-process. However, V is an element whose abundance is difficult to measure, therefore it is very little studied. \cite{prochaska2000} suggest that the isotope $^{51}$V is produced mainly by $\alpha$-process (SNe II), but is also produced by SNe Ia, which could explain our results.

\indent \cite{spina2015}, studying the open cluster Gamma Vel, found that some stars in this cluster present overabundance of some elements with respect to other stars of the cluster. They show that all of these overabundant elements are refractory elements and the most refractory (Sc) is the most overabundant element. They suggest  that  the  peculiar abundance pattern observed in some stars originated from the enrichment of the stellar atmospheres due to accretion of rocky objects, such as planets or planetesimals. Based in this and in the fact that the stars of the sample of \cite{adibekyan2012} and \cite{neves2009} are stars hosting planets, we believe the existence of some stars with peculiar chemical in the \cite{adibekyan2012} and \cite{neves2009} samples may be related to the accretion of planetary material.

\indent Our results indicate that the chemical abundances space is highly structured, such that only a subspace with dimensions smaller than total number of elements can be used to explain the bulk of the observed enrichment processes.

\section{Acknowledgements}

RBS is supported by PhD grant from CAPES -- the Brazilian Federal Agency for Support and Evaluation of
Graduate Education within the Education Ministry of Brazil. We thank Rodolfo Smiljanic for important contributions for the improvement of this paper. 

\bibliographystyle{mnras}
\bibliography{Referencia.bib}

\bsp
\label{lastpage}
\end{document}